\documentclass[10pt,preprint]{aastex}

\usepackage{savesym}
\usepackage{natbib}

\usepackage{amsmath,amssymb,amsthm,mathrsfs,dsfont}
\usepackage{multirow}
\usepackage{graphicx}
\usepackage{subfigure}
\usepackage{captcont}
\usepackage{float}
\usepackage{booktabs}
\usepackage{epstopdf}
\usepackage{color}
\usepackage{hyperref}

\def\be{\begin{equation}}
\def\ee{\end{equation}}
\def\kms{{\rm \,km\,s^{-1}}}
\def\s{{\rm \,s^{-1}}}
\def\Gyr{{\rm \,Gyr}}

\def\Mpc{{\rm \,Mpc}}
\def\kpc{{\rm \,kpc}}
\def\erg{{\rm \,erg}}
\def\keV{{\rm \,keV}}
\def\msun{{\,M_\odot}}
\def\dh70{\,h_{70}^{-1}}

\begin{document}

\title{Simulating the galaxy cluster ``El Gordo'' and
identifying the merger configuration}
\shorttitle{Simulating ACT-CL J0102--4915}
\shortauthors{Zhang, Yu, \& Lu}
\author{Congyao Zhang$^1$, Qingjuan Yu$^1$, and Youjun Lu$^2$}
\affil{
$^1$~Kavli Institute for Astronomy and Astrophysics, Peking
University, Beijing, 100871, China; yuqj@pku.edu.cn \\
$^2$~National Astronomical Observatories, Chinese Academy of
Sciences, Beijing, 100012, China
}

\begin{abstract}
The observational features of the massive galaxy cluster ``El Gordo''
(ACT-CL J0102--4915), such as the X-ray emission, the Sunyaev-Zel'dovich (SZ)
effect, and the surface mass density distribution, indicate that they
are caused by an exceptional ongoing high-speed collision of two galaxy
clusters, similar to the well-known Bullet Cluster. We
perform a series of hydrodynamical simulations to investigate the
merging scenario and identify the initial conditions for the collision in
ACT-CL J0102--4915. By surveying the parameter space of the various physical
quantities that describe the two colliding clusters, including their total
mass ($M$), mass ratio ($\xi$), gas fractions ($f_{\rm b}$),
initial relative velocity ($V$), and impact parameter ($P$),
we find out an off-axis merger with $P \sim 800 \dh70 \kpc$,
$V \sim 2500\kms$, $M \sim 3\times 10^{15}\msun$, and $\xi = 3.6$ that can
lead to most of the main observational features of ACT-CL J0102--4915.
Those features include the morphology of the X-ray emission with a remarkable
wake-like substructure trailing after the secondary cluster, the X-ray
luminosity and the temperature distributions, and also the SZ temperature
decrement. The initial relative velocity required for the merger is
extremely high and rare compared to that inferred from currently available
$\Lambda$ cold dark matter ($\Lambda$CDM)
cosmological simulations, which raises a potential challenge
to the $\Lambda$CDM model, in addition to the case of the Bullet Cluster.
\end{abstract}


\section{Introduction} \label{sec:introduction}

Clusters of galaxies are unique laboratories for exploring the nature
of dark matter (DM) and the structure formation in the universe
\citep[see a recent review by][]{Kravtsov2012}. In the $\Lambda$ cold dark
matter ($\Lambda$CDM)
cosmology, massive galaxy clusters are assembled via accretion and
mergers of galaxy groups or small clusters. Some clusters are
undergoing mergers and dynamically unrelaxed systems with distinctive
features, which are expected to provide deep insights not only into
the merging process but also into the physics of hierarchical
structure formation. For example, studies of the Bullet Cluster (1E
0657--56; e.g., \citealt{Markevitch2002}) have demonstrated almost
exclusively the collisionless nature of DM \citep{Clowe2004,
Clowe2006, Milosavljevic2007, Springel2007, Mastropietro2008}, and
also revealed that the relative velocity ($\sim 2700-4500\kms$)
required for the merger to form the Bullet Cluster may be too
high and rare to be compatible with the prediction from the
$\Lambda$CDM model and thus put a strong constraint on the model
\citep[e.g.,][but see \citealt{Watson2014,Thompson2015,LF15,Kraljic2015}]{Lee2010, Thompson2012,Bouillot2014}.

ACT-CL J0102--4915 (``El Gordo''), a Bullet Cluster-like cluster at a
redshift of $z=0.87$, was recently discovered by the Atacama
Cosmology Telescope (ACT) through its Sunyaev-Zel'dovich (SZ) effect
\citep{Marriage2011}.  ACT-CL J0102--4915 was also detected by the South
Pole Telescope (SPT) and Planck SZ surveys \citep{Williamson2011,
Planck2014}. Multi-frequency observational follow-ups, including those in the
optical, X-ray, infrared, and radio bands, have shown that ACT-CL J0102--4915
is a rare and exceptional system \citep{Menanteau2012, Jee2014, Lindner2014}
at least in the following points.
(1) It is the most massive X-ray and SZ bright cluster ($\sim 2-3\times
10^{15} \dh70\msun$) at $z \ga 0.6$ discovered so far.
(2) The offsets between its SZ and X-ray centroids ($\sim 600 \dh70\kpc$)
and between its SZ centroid and the mass center of its main cluster
component ($\sim 150 \dh70\kpc$) are quite large.
(3) The morphology of its X-ray emission is elongated with two extended
faint tails, possibly a ``wake''-like feature.
(4) It is currently the highest-redshift cluster that hosts a radio relic.
These observational features suggest that ACT-CL J0102--4915 is probably
undergoing a major merger with high relative velocity
($V \sim 1200-2300$, $2600$, or $2250\kms$ obtained in
\citealt{Menanteau2012}, \citealt{Donnert2014}, or \citealt{Molnar2015},
respectively). However, the probability is extremely low for the existence
of a massive major merger with such a high initial relative velocity in the
currently available large-volume cosmological simulations
(\citealt{Menanteau2012}; see also \citealt{Jee2014}), which raises a
potential challenge to the $\rm \Lambda CDM$ model, in addition to the
case of the Bullet Cluster.

To understand those distinctive observational
features of ACT-CL J0102--4915, it is important and necessary to investigate
its detailed merging behavior by performing $N$-body/hydro-numerical
simulations and reproducing the observations, which may further help to
constrain the $\rm \Lambda CDM$ model.

In this work, we perform a series of numerical simulations of mergers
of two massive clusters and find out the merger configurations
that can lead to a good match to various observations of ACT-CL
J0102--4915. Some simulations on collisions between two isolated
clusters have been carried out previously to investigate the nature
of some particular clusters \citep[e.g., the Bullet Cluster; see][]
{Springel2007, Mastropietro2008, Zuhone2009, Machado2013}.
For example, \citet{Molnar2015} have simulated cluster mergers by using
the FLASH code \citep{Fryxell2000} to reproduce the observations of
ACT-CL J0102--4915 (see also \citealt{Donnert2014}). They found that the
two extended faint X-ray tails of ACT-CL J0102--4915 may be reproduced
through a nearly head-on merger of two massive clusters, however, where
only nine sets of initial conditions of the merging configurations are explored.
Considering both the advantages and the disadvantages in different kinds of
hydrodynamical simulations, in this work we first perform a large number of
merger simulations ($\sim120$) by using the efficient GADGET-2 code.
We survey the parameter space and find out the configuration of
those mergers that can lead to a good match to various observations of
ACT-CL J0102--4915.
Then we further resimulate those mergers by using the FLASH code,
with which some substructures (shocks, eddies, etc.) can be more
accurately simulated.

The paper is organized as follows. In Section~\ref{sec:method},
we describe the method of simulating cluster mergers and generating
the mock observational maps. We present our simulation results in
Section \ref{sec:result}. Two types of the simulated merging systems
(i.e., the nearly head-on merger and the highly off-axis merger) are
explored.
Detailed comparison of the simulation results with the observational
ones is also presented in this section. We further discuss
the effects of the gas fraction profile of galaxy clusters on simulating
the ``El Gordo''. Conclusions are summarized in Section \ref{sec:conclusion}.

Throughout the paper, we assume a flat $\rm \Lambda CDM$ cosmology
model with $\Omega_{\rm m}=0.30$, $\Omega_{\rm \Lambda}=0.70$, and the
Hubble constant $H_0=70\,h_{70}\kms\Mpc^{-1}$.

\section{Method} \label{sec:method}

We simulate the mergers of galaxy clusters by adopting
the two types of publicly available numerical codes:
(1) the GADGET-2 code \citep{Springel2001} and (2) the FLASH code
\citep{Fryxell2000, Ricker2008}. The GADGET-2 code uses the smoothed
particle hydrodynamics (SPH) method to solve the gas hydrodynamics,
and it has advantages in computational speed and effective
resolution \citep{Springel2001}.  However, it may not handle shocks,
eddies, and fluid instabilities accurately \citep{Mitchell2009,Agertz2007}. The
FLASH code uses the piecewise-parabolic method (PPM; \citealt{Colella1984})
to solve the gas hydrodynamics and the adaptive mesh refinement (AMR)
method to reach high spatial resolution only where it is needed. The
FLASH code handles shocks, eddies and fluid instabilities better than
the SPH code, but it is time-consuming.

Considering the advantages of the GADGET-2 code in computational
speed (generally about one order of magnitude faster than the FLASH code for
the mergers with the same initial conditions in our simulations),
we first use the GADGET-2 code to survey the parameter space
for the initial conditions and configurations of cluster mergers and
search for the one that can ``best fit'' the observations of ACT-CL
J0102--4915 (denoted as the fiducial model(s)). The mass resolutions
of DM and baryonic gas are set to be $1.65 \times 10^9 \dh70 \msun$
and $1.68 \times 10^8 \dh70 \msun$, respectively.  We resimulate the
fiducial models (see Sections~\ref{sec:result:classA} and
\ref{sec:result:classB}) by using the FLASH code, which offers a
better treatment to the fine structures of the merging cluster.  The
box size of our simulations is $15.6\dh70\Mpc$ on each side, and the
finest resolution achieved is $7.6\dh70\kpc$.  In our simulations,
the two progenitor clusters of ACT-CL J0102--4915 are assumed to be
spherical halos, composed of collisionless DM and adiabatic
collisional gas; the shock heating is included, while the radiative
cooling and additive heating mechanisms
(e.g., active galactic nucleus (AGN) feedback) are
neglected.

The setup of the initial configuration of the merging cluster and the
method to analyze the simulation data are introduced in
Sections~\ref{sec:method:init} and \ref{sec:method:analysis} below,
respectively.

\subsection{Initial Configuration} \label{sec:method:init}

Considering a merger of two clusters, the masses of the primary
and the secondary clusters are denoted as $M_1$ and $M_2$ ($M_1\ge M_2$),
respectively, and the mass ratio is $\xi \equiv M_1/M_2
$.\footnote{Following the conventional use, the mass refers to the total mass within a radius
where the mean overdensity is $200$ times the critical density of the
universe at the cluster redshift, and the radius $r_{\rm 200}$ is
the corresponding radius.} Within the radius $r_{\rm 200}$,
the fractions of the baryon mass to the total mass are denoted as
$f_{\rm b1}$ and $f_{\rm b2}$ for the primary and the secondary
clusters, respectively. A Cartesian coordinate system $x'y'z'$ is
adopted for our simulations. The collision of those two clusters is
assumed to occur in the $x'-y'$ plane. The initial positions of the
cluster centers are set to be $(d_{\rm ini}/(1+\xi),\ P/(1+\xi),\ 0)$
and $( -d_{\rm ini}\xi/(1+\xi),\ -P\xi/(1+\xi),\ 0)$, respectively,
where $P$ is the impact parameter, and $d_{\rm ini}$ is twice the sum of
the radii $r_{200}$ of the two clusters.  The initial separation between
the two clusters is $\sqrt{d^2_{\rm ini} + P^2} \sim d_{\rm ini}$ as
$d_{\rm ini} \gg P$ for most cases studied in this work. The initial
velocities of the primary and the secondary clusters are set to be
$(-V/(1+\xi),\ 0,\ 0)$ and $(V\xi/(1+\xi),\ 0,\ 0)$, respectively,
with which the mass center of the merging system maintains at rest at
the origin, and $V$ is the initial relative velocity.

For each cluster, we assume that the DM density distribution follows
the Navarro-Frenk-White (NFW; \citealt{NFW97}) profile within $r_{\rm
200}$, i.e.,
\be
\rho_{\rm DM}(r) = \frac{\rho_{\rm s}}{r/r_{\rm s}(1+r/r_{\rm
s})^{2}}, \quad {\rm for\ } r \leq r_{\rm 200},
\label{eq:rhoDM1}
\ee
where $\rho_{\rm s}$ and $r_{\rm s}\equiv r_{\rm 200}/c_{\rm 200}$
are the scale density and radius, and $c_{\rm 200}$ is the
concentration parameter. For a cluster with a given mass,
$c_{\rm 200}$ can be obtained by using the mass-concentration
relation of \citet{Duffy2008}; and in this work we adopt the median
of the statistical relationship (see table 1 in \citealt{Duffy2008}). We assume an exponential truncation
of the DM density distribution outside $r_{\rm 200}$ to avoid a
divergent total mass.  More details can be found in
\citet{Kazantzidis2004}. The velocities of DM particles are
assigned according to the distribution function derived from the
Eddington's formula (eq.~4.46 in \citealt{Binney2008}).

The gas density distribution is assumed to follow the Burkert profile
\citep{Burkert1995},
\be
\rho_{\rm gas}(r)=\frac{\rho_{\rm c}}{(1+(r/r_{\rm c})^2)(1+r/r_{\rm
c})}, \quad {\rm for\ } r\le r_{\rm 200},
\label{eq:rhogas1}
\ee
where $r_{\rm c}$ is the core radius, which is typically half of
$r_{\rm s}$ \citep{Ricker2001}. The gas density profile outside of
$r_{\rm 200}$ is assumed to trace the density distribution of DM
(see eq.~4 in \citealt{ZYL14}). We set $r_{\rm c}=r_{\rm s}/2$ for the
primary cluster.  For the secondary cluster, we find that a smaller
$r_{\rm c}$ would provide a better fit to the observational X-ray
morphology according to our simulations. Therefore, we choose
$r_{\rm c}=r_{\rm s}/3$ for the secondary cluster. The normalization
factor $\rho_{\rm c}$ can be obtained by equating the baryon mass
fraction within $r_{\rm 200}$ to $f_{\rm b1}$ and $f_{\rm b2}$
for the primary and secondary cluster, respectively.
Assuming that the gas is in hydrostatic equilibrium and ideal (with a
heat capacity ratio of $\gamma=5/3$), the temperature and the specific
internal energy distribution of the gas in each progenitor cluster
can all be numerically determined \citep{Ricker2001}. The
effects of different gas density profiles are further discussed in Section~\ref{sec:result:fgas}.

We survey the parameter space for the initial configuration of the
merger, i.e., ($M_1$, $\xi$, $f_{\rm b1}$, $f_{\rm b2}$, $V$, $P$), in
order to find the parameter set(s) that can reproduce the observations
of ACT-CL J0102--4915. Some constraints and hints on those parameters
may be adopted according to other observations. For example, the masses
of the northwest (NW) and the southeast (SE) components of
ACT-CL J0102--4915 are estimated to be $M_1 = 1.6 \times 10^{15} \dh70\msun$
and $M_2 = 0.8 \times 10^{15} \dh70 \msun$, respectively, by the
weak-lensing observations, and the mass ratio $\xi = M_1/M_2 = 2$
\citep{Menanteau2012,Jee2014}. The median gas fraction within
$r_{\rm 200}$ of galaxy clusters is $\sim 0.13$, with a scatter of
$10\%$ to $20\%$ of the median value
\citep[][see also \citealt{Mantz2014}]{Battaglia2013}. According to
\citet{Menanteau2012}, the relative velocity of the two progenitor
clusters of ACT-CL J0102--4915 should be high ($\ga 1200\kms$).
The visually non-perfect symmetric configuration of ACT-CL J0102--4915
suggests that the collision should not be exactly head-on.
According to those constraints and hints, we explore the parameter
space of the merging clusters, summarized in Table~\ref{tab:ic_para}.
We run totally about 123 sets of parameters to find the best-fit
merging scenarios of ACT-CL J0102--4915.

\begin{table*}
\begin{center}
\caption{Initial merger parameters}
 \label{tab:ic_para}
\begin{tabular}{c|c|c|c|c}
  \hline
  \hline
 $M_1\ (10^{15}\msun)$ & $\xi$ & $V\ (\kms)$ & $P\ (\dh70\kpc)$ & $(f_{\rm b1},\,f_{\rm b2})$ \\
  \hline
  $1.3, 1.6, 2.0$ & $2$ & $1000, 2000, 3000, 4000$ & $50, 200, 400$ & (0.10,\,0.10) \\
  \hline
  $1.3, 1.4$ & $1.5, 2$ & $2500, 3000$ & $300$ & (0.10,\,0.10), (0.13,\,0.13) \\
  \hline
  $1.3$ & $2$ & $1500$ & $300$ & (0.10,\,0.10)  \\
  \hline
  $1.3$ & $4$ & $2500$ & $300$ & (0.10,\,0.10)  \\
  \hline
  $2.0, 2.5, 3.0$ & 2, 4 & $2000, 3000, 4000$ & $600, 800, 1000$ & (0.05,\,0.10)  \\
  \hline
  $2.2, 2.5$ & $3.6, 4$ & $2500$ & $800$ & (0.05,\,0.10), (0.06,\,0.12)  \\
  \hline
  $2.5$ & $3.6$ & $2500$ & $800$ & (0.08,\,0.10), (0.10,\,0.10)  \\
  \hline
  $2.5$ & $3.6$ & $500, 1500, 3500$ & $800$ & (0.05,\,0.10)  \\
  \hline
  $2.5$ & $5$ & $2500$ & $800$ & (0.05,\,0.10)  \\
  \hline
  $1.6$ & $3.6$ & $2500$ & $600$ & (0.05,\,0.10)  \\
  \hline
  \hline
\end{tabular}
\end{center}
\end{table*}

\subsection{Mocking Observations of Simulated Merging Systems}
\label{sec:method:analysis}

For any given snapshot of a simulated merging cluster, we can obtain
the projected maps of the mass surface density, the X-ray surface
brightness, and the thermal SZ emission in the observer's sky plane by
the following equations.
\begin{itemize}
\item The mass surface density at a position is given by
\be
\Sigma=\int_{\rm LOS}^{}(\rho_{\rm DM}+\rho_{\rm gas})d\ell.
\label{eq:density}
\ee
\item The X-ray surface brightness is given by
\be
S_{\rm X}=\frac{1}{4\pi (1+z)^4}\int_{\rm LOS}^{}n_{\rm
e}n_{\rm H}\Lambda(T_{\rm gas},\, Z) d\ell,
\label{eq:xray}
\ee
where $n_{\rm e}$ and $n_{\rm H}$ are the number densities of electron
and hydrogen, respectively; $\Lambda(T_{\rm gas},\, Z)$ is the cooling
function depending on gas temperature $T_{\rm gas}$ and metallicity $Z$.
We assume that the metal abundance of those simulated clusters is the
same as the typical one of clusters, i.e., $Z= 0.3 Z_{\odot}$, where
the solar metal abundance $Z_{\odot}$ is adopted from \citet{Anders1989}.
Consequently, we have $n_{\rm e}=1.2n_{\rm H}$ and $\rho_{\rm
gas}=1.4m_{\rm H}n_{\rm H}$, and $m_{\rm H}$ is the mass of a hydrogen
atom. The cooling function $\Lambda(T_{\rm gas},\, Z)$ is obtained by
using the MEKAL model in the XSPEC v12.8 package.\footnote{See
\href{http://heasarc.nasa.gov/xanadu/xspec/}{http://heasarc.nasa.gov/xanadu/xspec/}}
\item The change in the cosmic microwave background (CMB) temperature
at frequency $\nu$ by the thermal SZ effect is determined by
\begin{eqnarray}
\frac{\Delta T}{T_{\rm CMB}} & & =  \frac{\sigma_{\rm T}k_{\rm
B}}{m_{\rm e}c^{2}} \int_{\rm LOS}^{}n_{\rm e} T_{\rm gas} \times
\nonumber \\ & &
(Y_{0}+Y_{1}\Theta+Y_{2}\Theta^{2}+Y_{3}\Theta^{3}+Y_{4}\Theta^{4})d\ell,
\label{eq:sz}
\end{eqnarray}
where $T_{\rm CMB},\ \sigma_{\rm T},\ k_{\rm B},\ m_{\rm e}$, and $c$
represent the CMB temperature, the Thomson cross section, the
Boltzmann constant, the electron mass, and the speed of light,
respectively. In the above equation, $\Theta\equiv k_{\rm B}T_{\rm
gas}/m_{\rm e}c^{2}$, $Y_{0}=x_\nu\coth(x_\nu/2)-4$ is the
non-relativistic frequency function, $x_\nu=h\nu/k_{\rm B}T_{\rm
CMB}$, and $Y_{1}$, $Y_{2}$, $Y_{3}$, and $Y_{4}$ are the coefficients
for the polynomial approximation of the relativistic correction (see
eqs.~2.26--2.30 in \citealt{Itoh1998}). We set $\nu=150{\rm \,GHz}$ in
this study, and smooth the SZ map by a Gaussian kernel with the width
$\sigma_{\rm SZ}=270\dh70\kpc$ to simulate the ACT observation
(with an FWHM of $\sim 1.4'$). Considering the signal to noise ratio
(SNR $\sim 9$) of ACT-CL J0102--4915 \citep{Marriage2011}, the position
of the SZ centroid may have an error of ${\rm FWHM}/ {\rm SNR} \sim 70\dh70\kpc$.
\end{itemize}

The subscript ``LOS'' in the above equations~(\ref{eq:density})-(\ref{eq:sz})
indicates that the integrations are over the line of sight (LOS).
The LOS vector ($\hat{z}$) can be obtained by rotating
the reference vector ($\hat{z}' = (0, 0, 1)$) through
$\hat{z}=R_{\rm x''}(i)R_{\rm z'}(\alpha)\hat{z}'$, where
$R_{\rm x''}(i)$ and $R_{\rm z'}(\alpha)$ are the rotation matrices
about the $x''$-axis and the $z'$-axis by an angle of $i$ and $\alpha$,
respectively, and $\hat{x}''=R_{\rm z'}(\alpha)\hat{x}'$
($\hat{x}' = (1, 0, 0)$). In such a transformation, the value of the angle
between $\hat{z}$ and $\hat{z'}$ is $i$.

In order to make a thorough comparison with the X-ray observation of
ACT-CL J0102--4915 \citep{Menanteau2012}, we obtain the mock
\textit{Chandra} X-ray images of those simulated merging systems by
using the MARX software package in Section~\ref{sec:result:obs}.\footnote{MARX
is designed to perform detailed ray-tracing simulations
of \textit{Chandra} observations. See
\href{http://space.mit.edu/ASC/MARX/}{http://space.mit.edu/ASC/MARX/}}
The input X-ray surface brightness maps (see Eq.~\ref{eq:xray}) are
obtained from those simulations by using the FLASH code (see
Sections~\ref{sec:result:obs}). The energy range is
from $0.1$ to $12.0\keV$, with a resolution of $0.05\keV$. We adopt
the ACIS-I detector and the High Resolution Mirror Assembly (HRMA).
The exposure time of each observation is set to {\rm 60\,ks}, the same
as the observational one. The diffuse cosmic X-ray background (CXB) is
included in producing the mock data \citep[e.g.,][]{Hickox2006}
by assuming a power law of the total intensity of the CXB,
\be
I=I_0\left(\frac{E}{1\keV}\right)^{-\Gamma},
\label{eq:cxb}
\ee
where $I_0=10.9\,{\rm cnt\,cm^{-2}\s\keV^{-1}\,sr^{-1}}$ and
$\Gamma=1.4$. The photoelectric absorption by our Galaxy
along the line of sight is also taken into account. The hydrogen
column is assumed to be $0.03$ (in units of $10^{22}\,{\rm
atoms\,cm^{-2}}$) in the simulation. The mock data are reduced with
CIAO v4.6 tools.\footnote{See
\href{http://cxc.harvard.edu/ciao/}{http://cxc.harvard.edu/ciao/}} We
then perform spectral analysis of the mock data by using the absorbed
\verb"phabs*mekal" model in the XSPEC package.

Note here that we do not consider the detailed simulated SZ map in
this study. A detailed comparison of a simulated SZ map with the
observational one may be important in distinguishing models, if the
resolution of future SZ observations of ACT-CL J0102--4915 is
sufficiently high.

\section{Simulation Results}
\label{sec:result}

In this section, we present our simulation results and the constraints
obtained on the initial configuration of ACT-CL J0102--4915. We search
for the ``best-fit model'' among more than one hundred possible
merging cases by comparing them with observations. The best match is
identified based on the following criteria.
\begin{enumerate}
\item The projected distance between the mass density centers of the
primary and the secondary clusters is $\sim 700\dh70\kpc$ \citep{Jee2014}.
\item The positions of the centroids of the X-ray and the SZ emissions
and the distance between them ($\sim 600 \dh70\kpc$) are similar to
the observational ones (see Fig. 7 in \citealt{Jee2014}).
\item The morphology of the X-ray emission is similar to the
observational one (see Fig. 1 in \citealt{Menanteau2012}), and the
total X-ray luminosity $L_{\rm X} \simeq (2.19 \pm 0.11) \times
10^{45} {\,h_{70}^{-2}} \erg\s$ in the 0.5--2.0$\keV$ band.
\end{enumerate}

Two types of merger configurations are explored in our simulations performed by
 using the GADGET-2 code:
(1) nearly head-on (or low-$P$) mergers with small impact parameters,
i.e., $P \leq 500 \dh70\kpc$, comparable to the scale radius $r_{\rm s}$;
(2) highly off-axis (or high-$P$) mergers with large impact parameters,
i.e., $P \geq 500 \dh70\kpc$.
The former and the latter are denoted as case A and B mergers in this study, respectively. The behavior of case A and B mergers is detailed in
Sections~\ref{sec:result:classA} and \ref{sec:result:classB}, respectively.
And the resimulations of the fiducial models by the FLASH code and their
comparison with the observations are presented in Section~\ref{sec:result:obs}.
Furthermore, we discuss the effects of the gas fraction profile of galaxy
clusters on simulating the ``El Gordo'' in Section~\ref{sec:result:fgas}.

\subsection{Case A Mergers}
\label{sec:result:classA}

Case A mergers are nearly head-on collisions of two massive clusters,
which are extremely energetic events. In those merger events, the gas
component in the two progenitor clusters is shock-heated and strongly
disturbed due to the collision. By exploring the parameter space for
case A mergers, we find that a merger with the parameter set
$(M_1,\, \xi,\, f_{\rm b1},\, f_{\rm b2},\, V,\, P) = (1.3\times 10^{15}
\dh70\msun,\, 2,\, 0.10,\, 0.10,\, 3000\kms,\, 300 \dh70\kpc)$ can match
most of the observational features of ACT-CL J0102--4915, and the merger
model defined by this parameter set is denoted as fiducial model A
in the following text.

\begin{figure*} 
\centering
\includegraphics[width=1.0\textwidth]{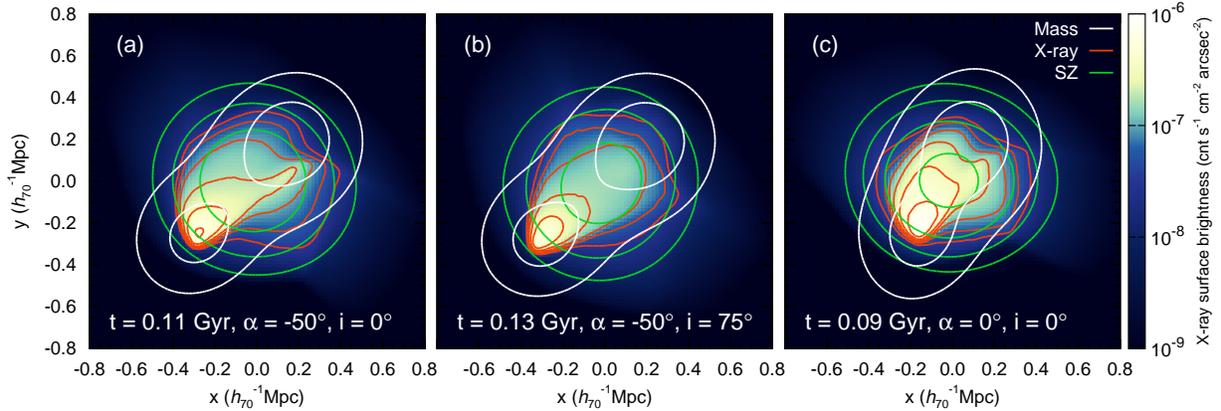}
\caption{X-ray surface brightness, mass surface density, and SZ effect
distributions for a merging cluster with the fiducial model A
configuration, simulated by using the SPH code. Panels (a) and (b)
show the results at different viewing directions, i.e.,
$(\alpha,\, i) =(-50\arcdeg,\, 0\arcdeg)$ and $(-50\arcdeg,\, 75\arcdeg)$,
at an evolution time of $t=0.11$ and $0.13\Gyr$ after passing the
pericenter, respectively. By these settings, the projected distance between
the two progenitor clusters is roughly the same ($\sim 700 \dh70 \kpc$).
Panel (c) shows the simulation results obtained from fiducial model
A at the viewing direction of $(\alpha,\, i) = (0\arcdeg,\, 0\arcdeg)$
at an evolution time of $t=0.09\Gyr$. The overlaid log-spaced contours
represent the projected mass surface density (white; the ratio between
two successive contour levels is 1.8), the X-ray surface brightness
(red, the ratio between two successive contour levels is 1.8) and the
SZ effect (green; the ratio between successive contour levels is 1.3).
Panel (b) shows the merging configuration in the case A mergers that
best matches to the observations (see Section~\ref{sec:result:classA}).
}
\label{pic:modelA_inclination}
\end{figure*}

Figure~\ref{pic:modelA_inclination} shows several snapshots of the
merger event, resulting from the SPH simulation (GADGET-2) of 
fiducial model A, viewing at a direction of $(\alpha,\, i) =
(-50\arcdeg,\, 0\arcdeg)$ (panel a), $(-50\arcdeg,\, 75\arcdeg)$ (panel
b), and $(0\arcdeg,\, 0\arcdeg)$ (panel c), at an evolution time of
$t=0.11$, $0.13$, and $0.09\Gyr$, respectively. (For simplicity, we set
the evolution time at the first pericentric passage as $t=0 \Gyr$.)
In each panel, the white, red, and green curves represent the contours
of the projected mass surface density, the X-ray surface brightness,
and the SZ effect, respectively.

In the first two panels, the viewing directions and the evolution
time are chosen so that the projected separation of the two clusters
is $\sim 700 \dh70\kpc$, similar to that of ACT-CL J0102--4915.
The two progenitor clusters just passed through and are running away
from each other. Besides the projected  distance, the morphology of
the X-ray surface brightness distribution of the simulated merging
cluster also depends on the evolution time and the viewing direction.
For example, in Figure~\ref{pic:modelA_inclination}, the X-ray
morphology is strongly asymmetric in panel (a), but not in panel (b).
Among the case A mergers that we simulate, fiducial model A
can generate an  X-ray surface brightness distribution similar to
the observational one, and its mass surface density distribution
is roughly consistent with that reconstructed by the weak-lensing
method \citep{Jee2014}, if the simulated merging cluster is viewed
at $t=0.13\Gyr$ and at a direction of
$(\alpha,\,i) \simeq (-50\arcdeg,\,75\arcdeg)$ (panel b).
In the X-ray morphology, the peak position of the X-ray surface
brightness  is close to the center of the secondary cluster after
the central gas core of the primary cluster is penetrated by the
secondary. More discussion on the peak positions of the X-ray and
the SZ maps will be presented in Section~\ref{sec:result:obs:peak}.
(For a general discussion of the positions of the X-ray and SZ
peaks and their separation, see \citealt{ZYL14, Molnar2012}.) Note that
compared to observations, few substructures are found in the
simulated merging cluster, which is probably due to the assumption
of a spherical symmetric initial configuration for the progenitor
clusters and the ignoring of galaxies in the progenitor clusters.

In panel (c), we can see a ``wake''-like X-ray structure in 
fiducial model A if viewing at the direction of
$(\alpha,\, i) = (0\arcdeg,\, 0\arcdeg)$ at $t=0.09\Gyr$. However,
at a  later evolution time, the ``wake'' becomes more asymmetric
(like the case shown in Fig.~\ref{pic:modelA_inclination}a). After
$t=0.15\Gyr$, a second peak emerges in the X-ray morphology,
located close to the center of the primary cluster. Usually the
merging process generates one tail (i.e., a matter stream connecting
the two merging clusters) and two wings (e.g., strong shocks in a
wing shape leading the secondary cluster) after the primary
pericentric passage. In Figure~\ref{pic:modelA_inclination}(c),
one of the wings is overlapping with the tail because of the non-zero
impact parameter, and the X-ray morphology
appears to be ``twin-tailed''. We find that the ``twin-tailed''
structure shown in panel (c) is obviously smaller and more
asymmetric than the observational one, and the (projected)
distance between the two clusters is $\sim 600 \dh70 \kpc$,
shorter than the constraint by the weak lensing. Therefore, we
conclude that panel (b) matches the observations better than other
cases in the case A mergers, although no ``twin-tailed'' structure
is produced in the X-ray morphology.

While the two clusters run away from each other, the wings become weaker
and weaker until they disappear. As the merger is off-axis,
the lifetimes of the two wings are different, and
therefore there is a time period in which only one wing and one tail
exist and the X-ray morphology also appears as ``twin-tailed''. This
is the case for the merging stage of ACT-CL J0102--4915 proposed in
\citet[][hereafter the MB model, i.e., $P=300\dh70\kpc$]{Molnar2015},
in which the ``twin-tailed'' morphology appears at a time
$t\sim480\,{\rm Myr}$ after the first core passage, much later than
that  shown in Figure~\ref{pic:modelA_inclination}(c). Compared with
the cases discussed in the MB model, the tails found in our simulation
are shorter as they emerge at an earlier merging stage. \citet{Molnar2015}
chose a larger concentration
parameter compared to the one adopted in our study,
and they found significant but narrow ``wake''-like structures
composed by one tail plus one strong wing. It appears, however,
that the weak wing does not completely disappear in their model
\citep[see Fig.~2 in][]{Molnar2015}. \citet{Donnert2014} also modeled
ACT-CL J0102--4915 but found no ``wake''-like structure, whose
simulations adopted a smaller impact parameter and a smaller gas
core for the secondary cluster.

\begin{figure*} 
\centering
\includegraphics[width=0.7\textwidth]{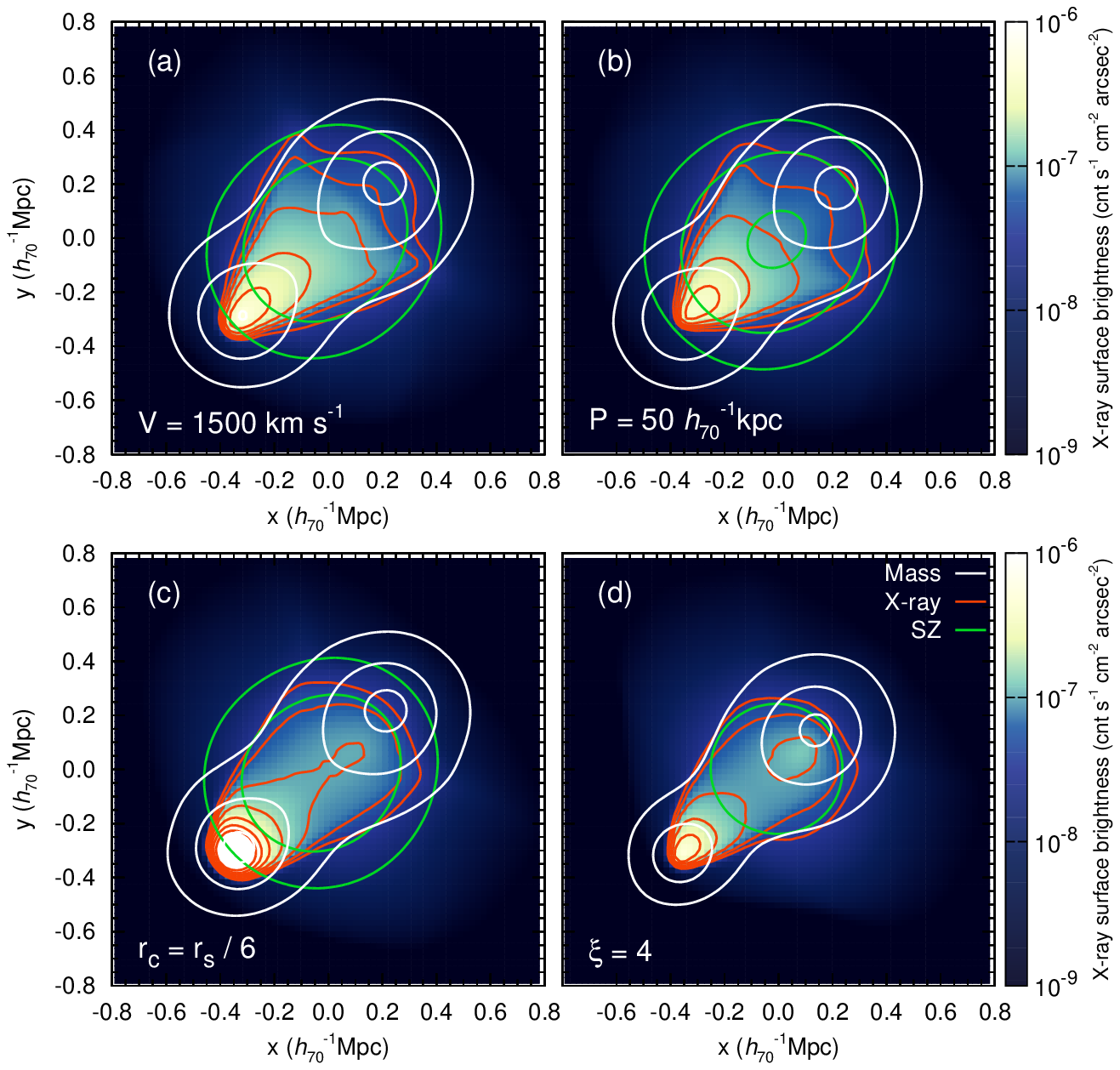}
\caption{Similar to those in Figure~\ref{pic:modelA_inclination},
but for the simulated merging clusters with initial configurations
slightly different from that of fiducial model A. Panels (a), (b),
(c), (d) show the results for the cases with only the relative velocity
$V$ ($=1500\kms$), the impact parameter $P$ ($=50\dh70\kpc$),
the core radius of the secondary cluster $r_{\rm c}$ ($=r_{\rm s}/6$),
or the mass ratio $\xi$ ($=4$) different from that for fiducial
model A, respectively. The viewing direction is set to be the same,
$(\alpha,\, i) =(-50\arcdeg,\, 75\arcdeg)$ for all the four cases
shown here. The snapshots shown in panels (a), (b), (c), (d) are at
a time $t=0.17$, $0.16$, $0.14$, and $0.14\Gyr$, respectively, in
order to keep the projected distance between the two clusters.
See Section~\ref{sec:result:classA}.
}
\label{pic:modelA_parameter}
\end{figure*}

Figure~\ref{pic:modelA_parameter} shows the simulation results of
several case A mergers, each with only one parameter, such as the
initial relative velocity (panel (a)), the impact parameter (panel (b)),
the core radius of the secondary cluster (panel (c)), or the mass
ratio (panel (d)), different from those of fiducial model A.
The snapshot time of each simulation shown in the figure is chosen
to keep the projected separation of the two clusters to be consistent
with the observation, as done above. This figure shows that the results
obtained from those different parameters do not match the observations
better than the result from fiducial model A. As seen from
Figure~\ref{pic:modelA_parameter}(a) and Figure~\ref{pic:modelA_inclination}(b),
the lower initial relative velocity results in a longer time required
for the interaction of the gas halos of the two progenitor clusters and
a longer time for the shocks to propagate farther away. Therefore,
the shocks (i.e., wing-like structure) driven by the collision appear
more significant in the X-ray map for the case of a merger with a lower
initial relative velocity than that with a higher velocity. The shape
of the X-ray emission in the central region thus tends to be more like
a triangle (or a bullet) in the case with a lower initial relative
velocity than that with a higher velocity. By comparing
Figure~\ref{pic:modelA_parameter}(b) with Figure~\ref{pic:modelA_inclination}(b),
we note that more distinct shocks can be formed through the merger with
a smaller impact parameter than that with a larger impact parameter,
as the collision with a smaller impact parameter is more violent.
By comparing Figure~\ref{pic:modelA_parameter}(c) with
Figure~\ref{pic:modelA_inclination}(b), we find that choosing a smaller
core radius for the secondary cluster may lead to an increase of the
X-ray emission in the center of the cluster; however, only one tail
structure tracing the secondary cluster is formed, which is inconsistent
with the observation. As seen from Figure~\ref{pic:modelA_parameter}(d)
and Figure~\ref{pic:modelA_inclination}(b), the merger with a smaller
secondary progenitor cluster (i.e., a large mass ratio $\xi$) is less
violent and may not be able to destroy the gas core of the primary
cluster; and in this case two peaks in the X-ray map may emerge, which
is also in contradiction with the X-ray observation of ACT-CL J0102--4915.

\subsection{Case B Mergers} \label{sec:result:classB}

Case B mergers are offset collisions of two massive clusters with
impact parameter $\ga 500\dh70\kpc$, their collision strengths are
less violent than those of the case A mergers. For case B mergers, the
gas cores of the primary clusters are not always destroyed after the
first pericentric passages; therefore, the merging systems may have
two peaks in their X-ray maps.  A single peak in the simulated X-ray
map as that for ACT-CL J0102--4915 may be also produced if the gas fraction
of the primary cluster is substantially lower than that of the secondary
cluster, e.g., $f_{\rm b1}= 0.05$, $f_{\rm b2} =0.10$; and the single
peak is close to the center of the secondary cluster.\footnote{As seen
in Section~\ref{sec:result:fgas}, a low gas fraction is
not necessary for the whole cluster. It is only needed in the central region
of the galaxy cluster, which is consistent with the current X-ray observations
\citep{Mantz2014}.} By exploring the parameter space, we find that a merger
with $(M_1,\, \xi,\, f_{\rm b1},\, f_{\rm b2},\, V,\, P) = (2.5\times
10^{15}\dh70\msun,\, 3.6,\, 0.05,\, 0.10,\, 2500\kms,\, 800\dh70\kpc)$
can match most of the observational features of ACT-CL J0102--4915, and
the merger model defined by this parameter set is denoted as fiducial
model B in the following text.

\begin{figure*} 
\centering
\includegraphics[width=1.0\textwidth]{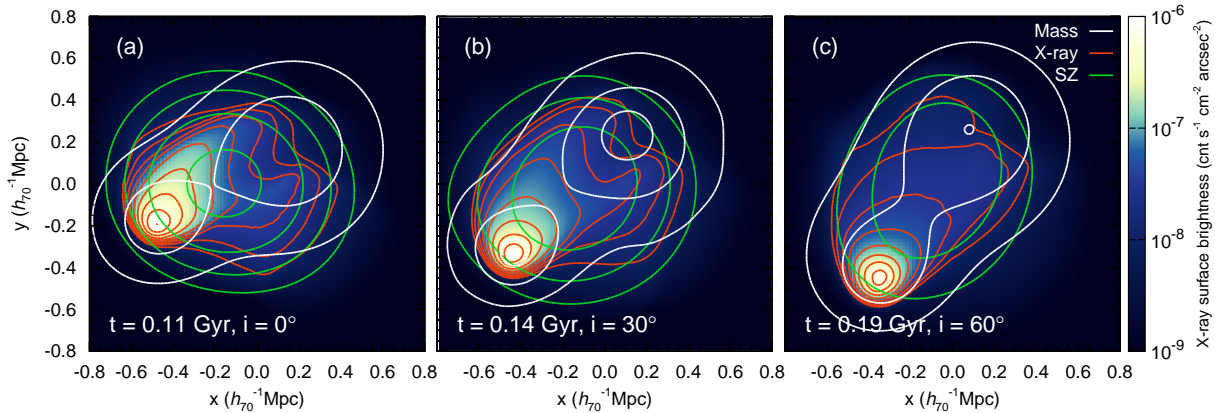}
\caption{Similar to those in
Figure~\ref{pic:modelA_inclination}, but for a merging cluster with
the fiducial model B configuration. The viewing direction is set to
$\alpha=-90\arcdeg$ and $i=0\arcdeg$, $30\arcdeg$, and $60\arcdeg$,
at the merging time of $t=0.11$ (panel (a)), $0.14$ (panel (b)),
and $0.19\Gyr$ (panel (c)), respectively.  This figure shows that
a wake (with two tails) trailing after the secondary cluster can be
produced by a merger with the fiducial model B configuration.
Panel (b) shows the merging configuration in the case B
mergers that best matches the observations (see Section~\ref{sec:result:classB}).
}
\label{pic:modelB_inclination}
\end{figure*}

Figure~\ref{pic:modelB_inclination} shows some snapshots of a merging
system resulting from fiducial model B, viewing at $(t,\,
\alpha,\, i)= (0.11\Gyr,\, -90\arcdeg,\, 0\arcdeg)$ (panel a),
$(0.14\Gyr,\, -90\arcdeg,\, 30\arcdeg)$ (panel b), and
$(0.19\Gyr,\, -90\arcdeg,\, 60\arcdeg)$ (panel c), respectively.
As seen from Figure~\ref{pic:modelB_inclination}, a ``wake'' clearly
exists trailing after the secondary cluster in the simulated X-ray
image, which is quite similar to the observational one of
ACT-CL J0102--4915 \citep{Menanteau2012}.  The ``wake'' structure
is more evident if viewing the merging system at the direction with
$i \sim 0\arcdeg-30\arcdeg$, which suggests that the merger event of
ACT-CL J0102--4915 should take place in a plane close to the sky plane,
consistent with the argument presented in \citet{Menanteau2012}.
The simulation results presented in Figure~\ref{pic:modelB_inclination}(b)
appear to match the observations the best. For example, the projected
distance between the centers of the two clusters in this simulation
is about $780\dh70\kpc$, roughly consistent with the observations;
the morphologies of the X-ray emission, the mass surface density,
and the SZ effect also match the observations well (see Fig.~1 in
\citealt{Menanteau2012}).

\begin{figure} 
\centering
\includegraphics[width=0.45\textwidth]{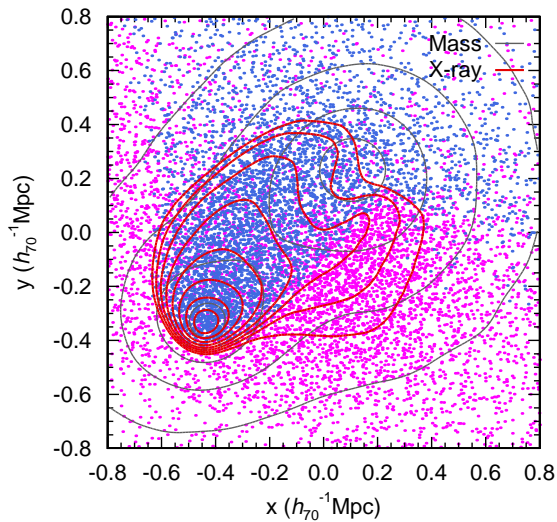}
\caption{Projected distribution of gas particles of a merging cluster
simulated by adopting the fiducial model B configuration and viewing
at the merging time $t=0.14\Gyr$ and at a direction of
($\alpha=-90\arcdeg,\, i=30\arcdeg$). The purple and blue points
represent the gas particles randomly selected from the primary
and the secondary clusters, respectively.  Red contours are the levels
of the X-ray surface brightness (log-spaced, $5.6\times10^{-7}$
to $10^{-8}\,{\rm cnt\,s^{-1}\,cm^{-2}\,arcsec^{-2}}$ from the
inner ones to the outer ones), and the grey ones are the mass
surface density (log-spaced, $0.56$ to $0.10\,{\rm g\,cm^{-2}}$
from the inner ones to the outer ones). This figure shows that
the top and the bottom parts of the wake are dominated by the
gas particles from the secondary and the primary clusters,
respectively. See Section~\ref{sec:result:classB}.
}
\label{pic:modelB_particles}
\end{figure}

We further investigate the gas distribution in the simulated merging
cluster resulting from fiducial model B, in order to understand
the origin of the wake shown in Figure~\ref{pic:modelB_inclination}.
Figure~\ref{pic:modelB_particles} shows the
projected distribution of gas particles for the snapshot shown in
Figure~\ref{pic:modelB_inclination}(b), where the purple
and the blue points represent some gas particles randomly selected
from those in the primary and the secondary clusters, respectively.
As seen from Figure~\ref{pic:modelB_particles}, the gas
core of the primary cluster is significantly displaced from the
center of its gravitational potential well (mainly determined by
the distribution of DM particles) because of the dissipative nature
of the gas collision; the distribution of gas particles originally
in the secondary cluster becomes elongated due to the compression
by the ram pressure from the primary cluster. The gas particles
from the primary and the secondary clusters have not effectively
mixed yet. Because of the large impact parameter of the case B
mergers, the resulting two wings in the X-ray morphology are not
as obvious as those resulting from the case A mergers. Gas particles
from the secondary and the primary clusters dominate the top and
the bottom parts of the wake, respectively, while the gas density
is low in the middle part of the wake. In this case, the wake
emerges mainly as a result of the specific overlapping positions of the
disturbed gas halos, and it is unlikely to be caused by the
merger-driven turbulence argued in
\citet[][see section~4.3 therein]{Menanteau2012}.

\begin{figure*}
\centering
\includegraphics[width=1.0\textwidth]{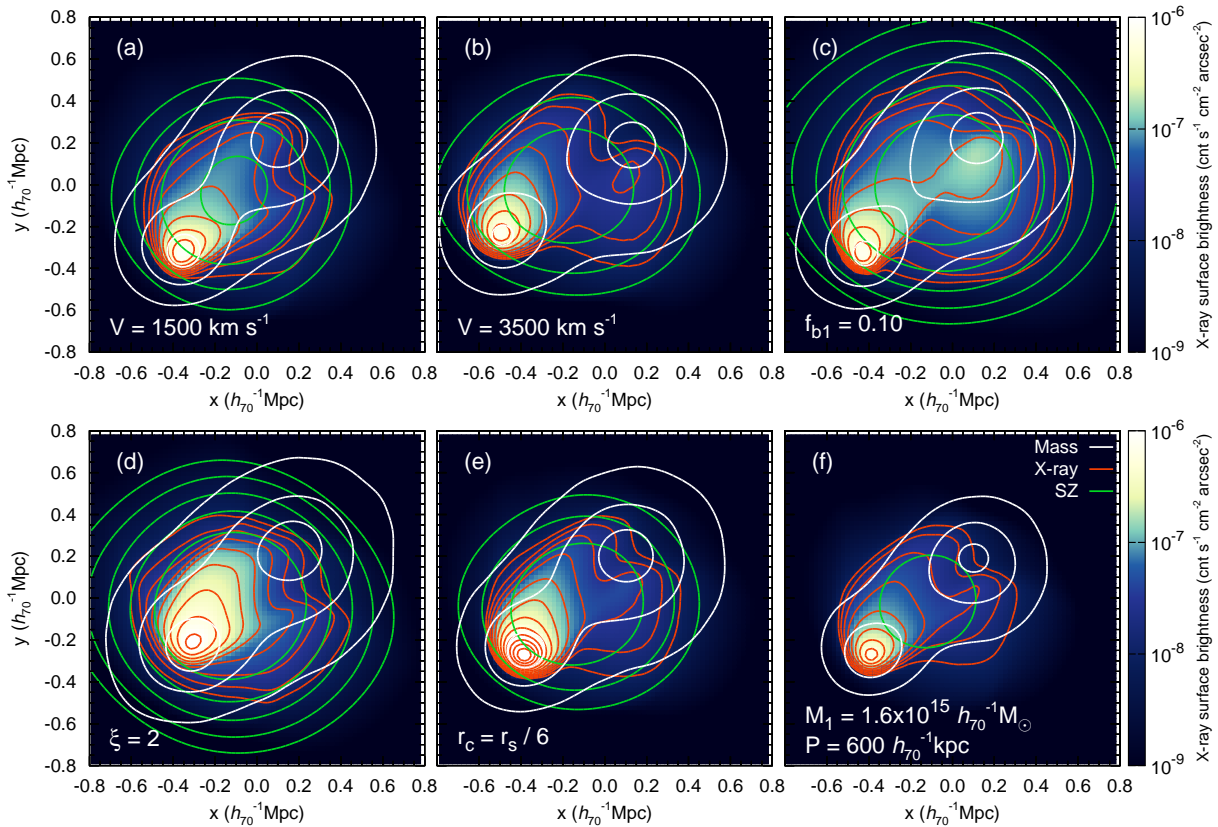}
\caption{Similar to that for Figure~\ref{pic:modelA_parameter},
but for simulated merging clusters with initial configurations
slightly different from that of fiducial model B. Panels (a)--(f) show the results for the cases with only the following parameters as labeled in each
panel different from those of fiducial model B: initial relative
velocity $V $ ($= 1500\kms$ and $3500\kms$), gas fraction of the primary
cluster $f_{\rm b1}$ ($=0.10$), mass ratio $\xi$ ($=2$), core radius of
the secondary cluster $r_{\rm c}$ ($=r_{\rm s}/6$), and mass $M_1$
($=1.6\times10^{15}\dh70\msun$) and impact parameter $P$ ($=600\dh70\kpc$).
The viewing direction is fixed at $(\alpha,\,i) = (90\arcdeg,\,-30\arcdeg)$.
The snapshots shown in panels (a)--(f) are at a time $t=0.15$, $0.11$, $0.14$, $0.11$, $0.12$, and $0.14\Gyr$, respectively, in order to keep the projected distance between the two clusters. Apparently, fiducial model B gives
a better match to the observations of ACT-CL J0102--4915, compared with
those cases shown in this figure. See Section~\ref{sec:result:classB}.
}
\label{pic:modelB_parameter}
\end{figure*}

Figure~\ref{pic:modelB_parameter} shows the simulation results on the
X-ray surface brightness, the mass surface density, and the SZ effect
distributions obtained from some of the case B mergers simulated in
this study, each with one or two parameters different from those of
fiducial model B. The snapshot time of each simulation shown in the
figure is chosen to keep the projected separation of the two clusters to
be consistent with the observation, as done above. As seen from the figure,
the results obtained from those different parameters do not match the
observations better than the result from fiducial model B. The
detailed effects of those different parameters on the resulting maps
are listed as follows.
\begin{itemize}
\item Panels (a) and (b) show a merger with only initial $V$
($=1500\kms$ and $3500\kms$, respectively) different from that
of fiducial model B. In the low-velocity case, the resulting
maps  appear to be similar to those of fiducial model B
(Fig.~\ref{pic:modelB_inclination}b). However, in the high-velocity
case, the bottom part of the wake becomes much stronger because
of the shorter interaction time between the two clusters. If an
even lower relative velocity is chosen, e.g., $V= 500\kms$,
there is only one tail trailing after the secondary cluster in
the resulting X-ray morphology. Therefore, $V\sim 1500-2500\kms$
is required in order to reproduce ACT-CL J0102--4915 in the case
B merger scenario, which is relatively lower compared with that
for fiducial model A.
\item
Panel (c) shows a merger with only $f_{\rm b1}$ ($=0.10$)
different from that of fiducial model B.  In this case, the gas
cores of the two progenitor clusters preserve themselves before the
secondary pericentric passage, and thus two peaks emerge in the
resulting X-ray map, which is apparently inconsistent with the X-ray
observation of ACT-CL J0102--4915.  In order to produce a single
X-ray peak by the case B merger scenario, the gas fraction of the
primary cluster must be lower than that of the secondary, and the
gas fraction of the secondary cluster should also not be too large
to form an unrealistic bright gas core in the center (e.g., $<0.13$).
\item
Panel (d) shows a merger with only the mass ratio $\xi$ ($=2$)
different from that of fiducial model B. Apparently,
the wake generated in this case is more asymmetric compared to that
resulting from fiducial model B ($\xi=3.6$). We also find that the
merger with a mass ratio of $5$, even smaller than that of fiducial
model B, however, results in a more asymmetric X-ray morphology as
well. According to those simulations, we conclude that a mass ratio of
$\xi \sim 3.6$ is preferred in order to re-produce the wake shown in
the X-ray map of ACT-CL J0102--4915.
\item
Panel (e) shows a merger with only the core radius of
the secondary cluster $r_{\rm c} (=r_{\rm s}/6)$ different from that
of fiducial model B. In this case, the resulting core of the X-ray
emission is brighter and the gradient of the X-ray emission is larger,
compared with those resulting from fiducial model B.
\item
Panel (f) shows a merger with only the primary
cluster mass ($M_1 = 1.6\times 10^{15}\dh70\msun$) and the impact
parameter ($P=600\dh70\kpc$) different from those of fiducial
model B.  Compared to fiducial model B, a smaller $P$ is adopted
here because of the smaller size of the adopted system. The shapes of
the X-ray surface brightness distribution and the SZ effect shown in
panel (f) are similar to those in
Figure~\ref{pic:modelB_inclination}(b); however, the total X-ray
luminosity resulting from this merging system is substantially smaller
than that from fiducial model B. In the case B scenario, a more
massive merging system (e.g., $M_1=2.5\times10^{15}\dh70\msun$) is
required in order to generate the total X-ray luminosity of ACT-CL
J0102--4915. Further discussion on the X-ray luminosity is detailed in
Section~\ref{sec:result:obs:temp}.
\end{itemize}

\subsection{FLASH Simulation Results and Comparison to Observations}
\label{sec:result:obs}

By surveying the parameter space of mergers of massive clusters,
we find that fiducial model B or A may be the solution to the
unique observational features of ACT-CL J0102--4915. The initial
conditions are summarized in Table~\ref{tab:results}. For a detailed
comparison, we re-run the simulations for fiducial models A and B
by using the FLASH code, respectively, because the FLASH code handles
shocks better than the GADGET-2 code. Figure~\ref{pic:flash_model}
shows the results obtained from the FLASH simulations of fiducial
models A (left panel) and B (right panel), respectively. We find that
the main merging structures obtained from the FLASH simulations are
well consistent with those obtained from the GADGET-2 simulations
(see Figs.~\ref{pic:modelA_inclination}b and \ref{pic:modelB_inclination}b
for comparison; for example, the difference between the amplitudes of
the X-ray peaks obtained from the two codes is not more than $5\%$.),
except that the shock structures resulting from the FLASH simulations
are sharper. This consistence supports the robustness of our method,
i.e., first surveying the parameter space of cluster mergers and
singling out the parameter set(s) that can lead to a close match to
the observations of ACT-CL J0102--4915 through efficient GADGET-2
simulations, and then mining out the details of the singled-out
mergers by doing the FLASH simulations. Below we present the
comparison between the FLASH simulation results and the observations
of ACT-CL J0102--4915 in several different aspects, i.e., the X-ray
surface brightness distribution, the positions of the centroids of
the X-ray emission and the SZ effect, the total X-ray luminosity
and the temperature distributions of electrons in the merging cluster,
the Mach number crossing the shock discontinuity, and the relative
radial velocity between the NW and the SW components of the cluster,
respectively. The main results are summarized in Table~\ref{tab:results}.

\begin{deluxetable}{cccccc}
\tabletypesize{\footnotesize}
\tablecaption{Summary of the initial conditions and the cluster properties
in the models and observations}
\startdata
\\ \hline \hline
  \multicolumn{6}{c}{Initial conditions} \\ \hline
   & Model A & Model B & Extended Model B & MB model &  \\ \hline
  $M_1\ (10^{15}\msun)$ & $1.3$ & $2.5$ & $2.5$ & $1.4$ &  \\\hline
  $\xi$ & $2.0$ & $3.6$ & $3.6$ & $1.9$ &  \\\hline
  $(f_{\rm b1},\,f_{\rm b2})$\tablenotemark{1}
& ($0.10,\,0.10$) & ($0.05,\,0.10)$ & ($0.11,\,0.12)$ & $(0.14,\,0.14)$ &  \\ \hline
  $V\ (\kms)$ & $3000$ & $2500$ & $2500$ & $2250$ &  \\ \hline
  $P\ (\dh70\kpc)$ & $300$ & $800$ & $800$ & $300$ &  \\ \hline
  Gas density \tablenotemark{2}
& \multirow{2}{*}{Burkert} & \multirow{2}{*}{Burkert} &
  \multirow{2}{*}{Power law} & \multirow{2}{*}{non-isothermal/Burkert} & \\
  profile & & & & \\
  \hline
  \hline
  \multicolumn{6}{c}{Measurements in the models and observations} \\ \hline
  & Model A & Model B & Extended model B & MB model \tablenotemark{3}
& Observation \\\hline
  $t\ (\Gyr)$ \tablenotemark{4}
& $0.13$ & $0.14$ & $0.17$ & $0.42$ & -- \\\hline
  $(\alpha,\,i)$ \tablenotemark{5}
& ($-50\arcdeg,\,75\arcdeg$) & ($-90\arcdeg,\,30\arcdeg$) & ($-90\arcdeg,\,40\arcdeg$) & ($-90\arcdeg,\,55\arcdeg$) & -- \\\hline
  $d_{\rm m}\ (\dh70\kpc)$ \tablenotemark{6}
& $740$ & $780$ & $890$ & $930$ & $\sim700$  \\\hline
  $d_{\rm SZ-X}\ (\dh70\kpc)$ \tablenotemark{7}
& $400$ & $440$ & $570$ & $790$ & $\sim600$  \\\hline
  Wake-like \tablenotemark{8}
& \multirow{2}{*}{No} & Yes &
  \multirow{2}{*}{Yes} & \multirow{2}{*}{Yes} & \multirow{2}{*}{Yes}\\
  structure & & (the best match) & & & \\\hline
  $\delta T_0\ ({\rm \mu K})$ \tablenotemark{9}
  & $-1430$ & $-1130$ & $-1850$ & $-1030$ & $-1046\pm116$  \\\hline
  $L_{\rm X}$ \tablenotemark{10}
& \multirow{2}{*}{$2.48 \pm 0.03$} & \multirow{2}{*}{$2.05 \pm 0.03$} & \multirow{2}{*}{$2.08 \pm 0.03$} & \multirow{2}{*}{$1.77 \pm 0.13$} & \multirow{2}{*}{$2.19 \pm 0.11$} \\
  $(10^{45} {\,h_{70}^{-2}}\erg\s)$ & & & & \\\hline
  $T_{\rm X}\ (\keV)$ \tablenotemark{11}
 & $15.8\pm1.2$ & $15.0\pm1.3$ &  $18.0\pm1.8$ & $9.9\pm0.6$ &
  $14.5\pm1.0$ \\\hline
  $\mathcal{M}$ (SE,\,NW) \tablenotemark{12}
& $(2.9,\,2.4)$ & $(2.7,\,2.5)$ & $(2.4,\,1.5)$ & $(4.6,\,-)$ & $(-,\,2.5^{+0.7}_{-0.3})$ \\\hline
   \multirow{2}{*}{$V_{\rm r}\ (\kms)$} \tablenotemark{13}
& $960$ & $1820$ & $2060$ & $640$ & $-$ \\
  & $(560)$ & $(910)$ & $(1200)$ & $(590)$ &
  $(586 \pm 96$ / $731 \pm 66)$\tablenotemark{14}
\\\hline
  X-ray extension \tablenotemark{15}
& No & No & Yes & Yes & Yes \\\hline
\enddata
\tablecomments{~Lists of the initial conditions for different models
(i.e., fiducial model A, fiducial model B, extended model B, and MB model)
and comparison of the measurements between the models  and the observations
\citep{Menanteau2012, Jee2014, Lindner2014}.}
\tablenotetext{1}{Gas fractions of the primary ($f_{\rm b1}$) and the secondary ($f_{\rm b2}$) clusters at the radius $r_{200}$.}
\tablenotetext{2}{Type of the gas density profile of the primary
  cluster adopted in the simulations, i.e., ``Burkert'': the gas density
  profile is assumed to follow the Burkert profile (see Eq.~\ref{eq:rhogas1});
  ``Power law'': the gas density profile is set by assuming that the cumulative
  gas fraction profile follows a power-law form (see Eq.~\ref{eq:fgas}).
  Note that the gas density profile used in
\citet{Molnar2015} is the non-isothermal $\beta$ model with $\beta=1$ (see their
eq. 2), and we also label the MB model as ``Burkert'', since the Burkert and the
non-isothermal $\beta(=1)$ models follow the same tendency at both large radii
(proportional to $r^{-3}$) and small radii.}
\tablenotetext{3}{The values listed for the MB model are measured from our simulation results obtained by using the initial conditions of the MB models.}
\tablenotetext{4}{Evolution time of the merging system.}
\tablenotetext{5}{Viewing direction.}
\tablenotetext{6}{The projected distance between the
mass density centers of the primary and the secondary clusters.}
\tablenotetext{7}{Offset between the SZ and the X-ray
  centroids (see Section~\ref{sec:result:obs:peak}).}
\tablenotetext{8}{Existence of the wake-like structure in the X-ray
  image of the merging system. Model B apparently provides
  the best match to the wake-like structure in the observation.}
\tablenotetext{9}{Central temperature decrement
  of the SZ effect (see Section~\ref{sec:result:obs:xray}). The
  nonthermal pressure is not considered in the study, which may lead to
  the overestimation of the central temperature decrement by dozens of
  percent in the models.}
\tablenotetext{10}{Total X-ray luminosity in the $0.5-2.0\keV$ band
  (see Section~\ref{sec:result:obs:temp}).}
\tablenotetext{11}{Best-fitted X-ray temperature (see Section~\ref{sec:result:obs:temp}).}
\tablenotetext{12}{The Mach numbers, derived from
  the spectroscopic-like temperature profile across the SE and
  the NW shocks (see Section~\ref{sec:result:obs:mach}).
  (No clear NW shock is observed in the MB model.)}
\tablenotetext{13}{ Relative radial velocity between the NW and the SE components of the cluster,
   measured by two different methods.
   The top row presents the values directly estimated from the
   peculiar velocities of the NW and the SE mass centers in the simulations;
   the bottom row presents the measurements from the radial velocity
   distributions of the DM particles (models) and the galaxies (observation)
   along the LOS. It is important to note that the values obtained in the
   latter method (see also fig.~9 in \citet{Menanteau2012})
   may be significantly lower than the relative radial velocity of the mass centers (see discussions in Section~\ref{sec:result:obs:vel}).}
\tablenotetext{14}{The observed relative radial
  velocity along the LOS between the NW and the SE cluster components
  ($586 \pm 96\kms$), and between the NW component and the brightest cluster
galaxy (BCG) located
  in the SE component ($731 \pm 66\kms$) \citep{Menanteau2012}.}
\tablenotetext{15}{Extension of the X-ray emission in the
  outer region of the merging cluster (see Section~\ref{sec:result:fgas}).}
\label{tab:results}
\label{tab:results}
\end{deluxetable}

\subsubsection{Morphology of the X-Ray Surface Brightness and SZ Temperature
Decrement}
\label{sec:result:obs:xray}

\begin{figure*} 
\centering
\includegraphics[width=0.8\textwidth]{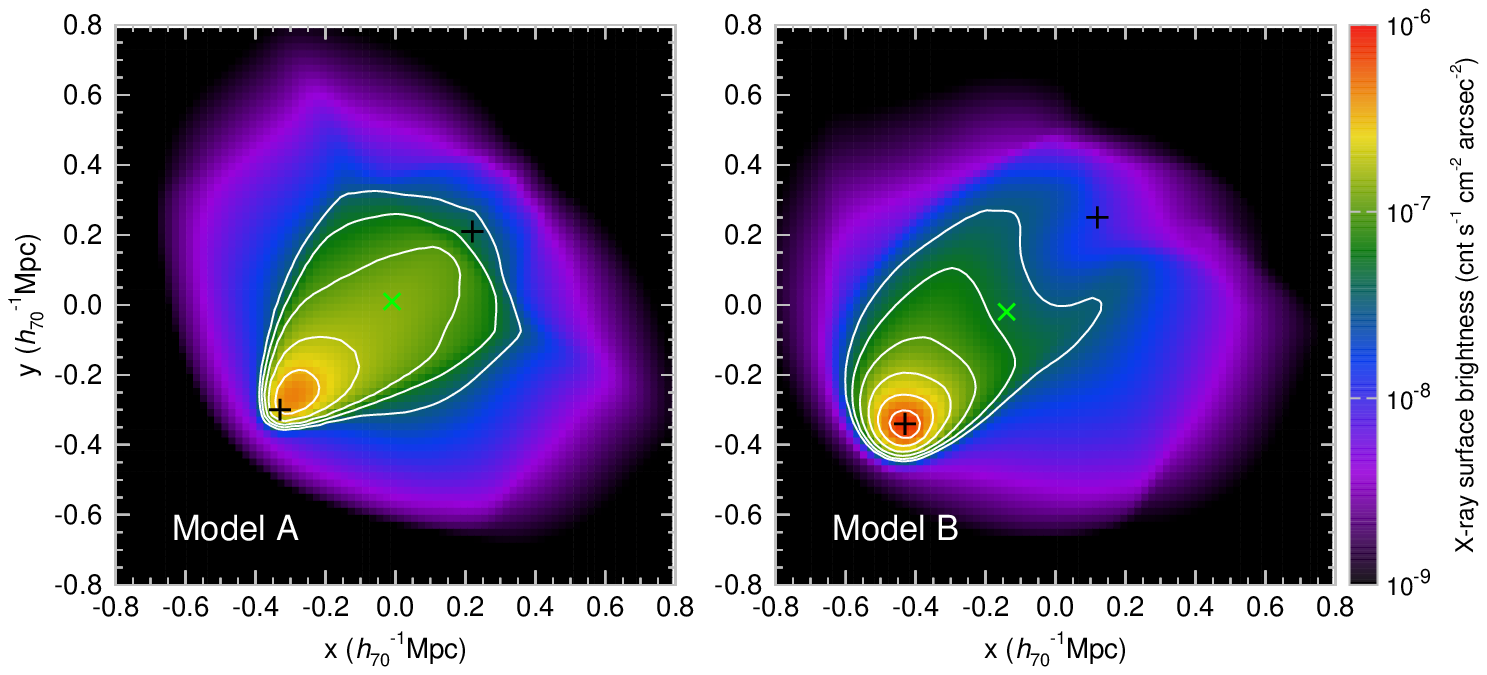}
\caption{X-ray surface brightness distributions obtained from the
FLASH simulations. Left and right panels show the results for
fiducial models A and B, respectively. In each panel, the symbols `+'
(black) and `$\times$' (green) symbols mark the positions of the mass
centers of the two clusters and the centroid  of the SZ map,
respectively. The white log-spaced contours represent the levels of
X-ray brightness of $3.2\times10^{-8}$, $5.6\times10^{-8}$, $10^{-7}$,
$1.8\times10^{-7}$, $3.2\times10^{-7}$, and $5.6\times10^{-7}$
from outside to inside, respectively. The morphological
features of fiducial Model A and B are summarized in
Table~\ref{tab:results}. See also Section~\ref{sec:result:obs:xray}.
}
\label{pic:flash_model}
\end{figure*}

As seen from the left panel of Figure~\ref{pic:flash_model}, we find the
following points for fiducial model A. (1) The simulated morphology
of the X-ray emission has a cometary appearance, which is consistent with
the \textit{Chandra} X-ray image of ACT-CL J0102--4915 \citep{Menanteau2012}.
However, our simulation cannot produce a wake-like feature trailing after
the secondary cluster as seen in ACT-CL J0102--4915.
(2) The core of the simulated X-ray emission is not as bright as the
observational one, because the gas in the core of the secondary cluster
is partly stripped off due to the nearly head-on collision, which leads
to a fainter X-ray core. Setting a smaller core radius $r_{\rm c}$
for the secondary cluster (Eq.~\ref{eq:rhogas1}), may reduce this
inconsistency but result in some other inconsistency
(see Figs.~\ref{pic:modelA_parameter}c and \ref{pic:modelB_parameter}e).

As seen from the right panel of Figure~\ref{pic:flash_model}, we find
the following points for fiducial model B. (1) The gas core of the
smaller cluster survives after the first pericentric passage. The X-ray
emission core is slightly brighter than the observational one.
(2) A remarkable wake-like feature, similar to the observation, is
reproduced. Figure~\ref{pic:flash_1d} shows the comparison between
the X-ray surface brightness across the wake resulting from
fiducial model B and that from the \textit{Chandra} observation.
As seen from this figure, the simulation results can well match the
twin-tailed structure found by the \textit{Chandra} observation.
(3) The morphology of the X-ray emission in the inner region (see Fig.~\ref{pic:flash_specT}, the inner regions~1, 2, and 3) of the
merging system is roughly the same as the observations.

For both fiducial models A and B, the resulting X-ray surface
brightness distribution rapidly decreases to an extremely low level at
the outer region (regions 4 and 5 marked in Figs.~\ref{pic:flash_specT}b1
and \ref{pic:flash_specT}c1) of the merging system. However, the X-ray
emission of ACT-CL J0102--4915 extends to a relatively large scale
($\sim 1\dh70\Mpc$) and the decrease of the surface brightness is not
as steep as the simulation ones at the outer region
(see a careful comparison shown in Fig.~\ref{pic:flash_specT}).
We note here that (1) considering the contamination from
the CXB cannot solve this discrepancy; (2) increasing the total mass
of the merging system does not lead to a better match to the X-ray
morphology, especially for the extended X-ray emission in the outer
region; and (3) choosing a snapshot at a later merging time does not
lead to a significant improvement in matching the X-ray morphology
(cf.\ the MB model), either. In our simulations, a given gas fraction normalized
at the radius $r_{200}$ is adopted for the gas density distribution, which
might not represent the real distribution well (see
Fig.~\ref{pic:modelB_e_profile}). We find that setting a cumulative gas
fraction as a function of the radius for the primary cluster
($f_{\rm b1}\sim0.05$ at $0.1r_{\rm 200}$ and $f_{\rm b1}\sim0.11$ at
$r_{\rm 200}$, motivated by the cosmological simulations and the X-ray
observations; \citealt{Battaglia2013,Mantz2014}) could either solve the above
discrepancy or provide a natural explanation to the low gas fraction of the
primary cluster required in fiducial model B (denoted as extended
model B). More discussions are in Section~\ref{sec:result:fgas}.

Because of the limited angular resolution of the SZ observation
(i.e., $1.4'$ of ACT at {\rm 148\,GHz}), we focus on the central
temperature decrement of the SZ effect but not the morphology. The
strength of the SZ signal at the center for fiducial models A
and B is $-1430$ and $-1130\,{\rm \mu K}$, respectively. The result
from model B is in agreement with the measured temperature
decrement of ACT-CL J0102--4915, i.e., $-1046\pm116\,{\rm \mu K}$, in \citet{Marriage2011}. However, the result from fiducial model A is
about $30\%$ larger. It is worth noting that the
non-thermal pressure, which is not considered in this study, may have a
non-negligible effect on modeling the SZ emission \citep{Battaglia2012}, and
thus the central temperature decrement of the SZ effect obtained in the models
may be overestimated by dozens of percent
\citep{Trac2011, Bode2012}.

\subsubsection{Peak Positions of the X-Ray and SZ Emissions}
\label{sec:result:obs:peak}

Positions of the peaks of the X-ray emission and the SZ effect of
merging clusters contain abundant information about the merging
process. Different dependences of the X-ray and the SZ signals on the
gas density and temperature distributions may induce a significant
SZ-X-ray offset in a massive merging system after the first
pericentric passage, when the X-ray peak locates near the center of
the secondary cluster (i.e., the ``jump effect'', see the
demonstration in \citealt{ZYL14}) and the SZ peak locates close to the
center of the primary cluster. ACT-CL J0102--4915 is a typical
example, of which the SZ-X-ray offset is about $600\dh70\kpc$
\citep{Menanteau2012}.

The SZ-X-ray offsets obtained from fiducial models A and B are
both close to $400 \dh70 \kpc$, somewhat smaller than that given
by observations (i.e., $600\dh70 \kpc$). The SZ centroids resulting
from the models are separated from the centers of the primary
clusters by $280 \dh70 \kpc$, which are somewhat larger than that
of ACT-CL J0102--4915 reported in \citet[][$\sim150\dh70\kpc$]{Jee2014}.
Considering the low angular resolution of the SZ observation,
the uncertainty in the SZ centroid estimate is $\sigma_{\rm peak}\sim
1.4'/{\rm SNR}\sim 70\dh70\kpc$. The differences between the model
results and the observations on the SZ-X-ray offset or the separation
between the SZ centroid and the mass center of the primary cluster
are about the same as the uncertainty, which suggests that our model
results are roughly consistent ($\sim2\sigma_{\rm peak}$) with the observations
on these aspects.

For ACT-CL J0102--4915, \citet{Jee2014} found that the distance between
the centroid of the X-ray emission and the center of the secondary
cluster is $\sim 60 \dh70 \kpc$; and the X-ray centroid leads the mass
surface density peak of ACT-CL J0102--4915 if the merging cluster is
viewed soon after the first core passage. The spatial offsets between
the X-ray centroid and the secondary cluster center of the simulated
merging clusters are $\sim 50 \dh70 \kpc$ in fiducial model A and
$\sim 5 \dh70 \kpc$ in the fiducial model B, respectively. However,
it appears that the X-ray centroid resulting from model A or 
model B does not lead the mass surface density peak in the direction
as shown in the observation. Furthermore, we do not find any case
whose X-ray centroid leads the mass surface density peak by more than
$50 \dh70 \kpc$ among the simulated merging clusters. We further examine the
two scenarios suggested in \citet{Jee2014}, i.e., (1) the merger is viewed
before the first apocentric passage, and has a low initial merger speed;
(2) the merger is viewed after the first apocentric passage, and has a high
initial merger speed; and we find that neither
of them can be a good solution because of the mismatch between the
simulated X-ray morphology and the observational one.

If viewing SZ emissions with a substantially higher angular resolution
(i.e., $\sigma_{\rm SZ}=10\dh70\kpc$), we may see two peaks in the SZ map
in fiducial model B (also in extended model B; see Section~\ref{sec:result:fgas}). The primary one is near the center of
the primary cluster, and the secondary one is close to the center of the
secondary cluster. However, no secondary SZ peak exists in the
high-resolution SZ image of fiducial model A. Therefore, the future
SZ observations with the detailed substructures of ACT-CL J0102--4915
may provide more constraints on the merging scenarios.

\subsubsection{X-Ray Luminosity and Temperature Distributions}
\label{sec:result:obs:temp}

\begin{figure*} 
\centering
\includegraphics[width=0.45\textwidth]{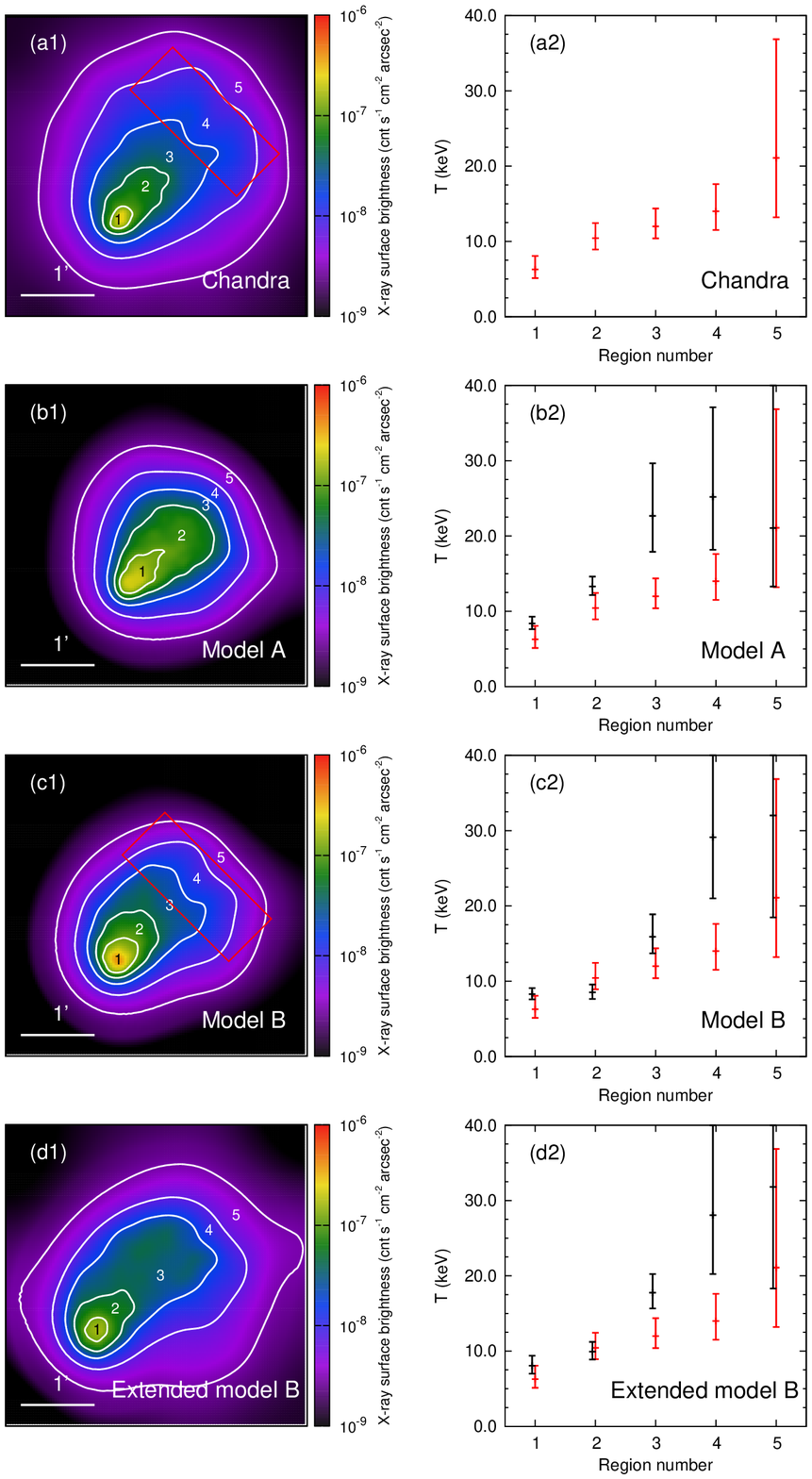}
\caption{\textit{Left panels:} image of the \textit{Chandra} X-ray
emission (panel (a1)), mock \textit{Chandra} X-ray images resulting
from fiducial model A (panel (b1)), fiducial model B (panel(c1))
and extended model B (panel (d1)). The contour levels are
$0.29\times10^{-8},\ 0.70\times10^{-8},\ 1.8\times10^{-8},\ 4.5\times10^{-8}$,
and $1.1\times10^{-7}\,{\rm cnt\,s^{-1}\,cm^{-2}\,arcsec^{-2}}$ from
outside to inside, respectively. The intersections of the X-ray surface brightness
distribution across the wake region in the red boxes are shown in
Figure~\ref{pic:flash_1d} (see Section~\ref{sec:result:obs:xray}).
\textit{Right panels:} temperatures estimated from the spectra of
X-ray emission from different extraction regions corresponding to the
region number marked in the left panels
(see Section~\ref{sec:result:obs:temp}). The red points are the result
for the observation; the black points are those for the models, which
are slightly shifted to the left to show a clear comparison with the
observation. The maximums of the X-ray surface brightness in panels (b1)
and (c1) are lower than those shown in Figure~\ref{pic:flash_model},
because the exposure time of the mock \textit{Chandra} X-ray images is
limited and a larger smoothing scale is adopted in the images.
The mock X-ray image and the temperature distribution of extended model
B are similar to those of fiducial model B; but the X-ray emission in
the outer region obtained from extended model B
is stronger, which is comparable with that of the observation
(see Section~\ref{sec:result:fgas}).
}
\label{pic:flash_specT}
\end{figure*}

\begin{figure} 
\centering
\includegraphics[width=0.45\textwidth]{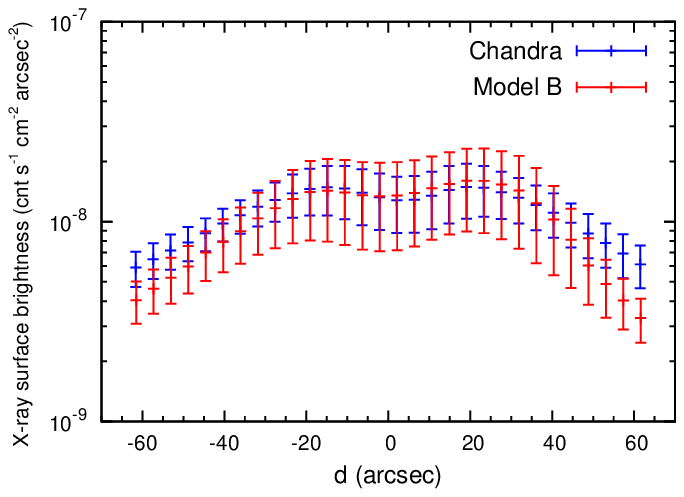}
\caption{
X-ray surface brightness distribution across the wake of
ACT-CL J0102--4915 from the red boxes in Figure~\ref{pic:flash_specT}.
The twin-tailed structure resulting from fiducial model B (red)
can well match that of the \textit{Chandra} observation (blue).
See Section~\ref{sec:result:obs:xray}.
}
\label{pic:flash_1d}
\end{figure}

We obtain the mock \textit{Chandra} X-ray images of merging clusters
($0.5-2.0\keV$) by considering the exposure correction and the adaptive
kernel smoothing. The left panels of Figure~\ref{pic:flash_specT} show
the \textit{Chandra} observation (panel (a1)) and the mock X-ray images
obtained from fiducial models A (panel (b1)) and B (panel (c1))
, respectively, for which the original X-ray images are shown in
Figure~\ref{pic:flash_model}. As seen from the figure, the
substructures (e.g., shocks, wake) in the mock images appear less
sharp than those in the original images (see Fig.~\ref{pic:flash_model})
because of the adopted smoothing over a large scale to mimic the
\textit{Chandra} observations.  Compared to the observations, both
models result in a more concentrated X-ray-emitting gas distribution
as discussed in Section~\ref{sec:result:obs:xray}. Panel (d1) presents
the results of extended model B (see Section~\ref{sec:result:fgas}).

We extract the mock \textit{Chandra} spectrum from the MARX simulated
images for both models, where the extraction region is set to those
areas between the innermost and the outermost contours shown in each
of the left panels of Figure~\ref{pic:flash_specT}, similar to the
analysis performed for the X-ray observation of ACT-CL J0102--4915 in
\citet{Menanteau2012}.  We fit the spectra by using the
\verb"phabs*mekal" model, and obtain the mean temperature of the
merging system resulting from fiducial model A or B as
$ T_{\rm X} = (15.8\pm1.2)\keV$ or $(15.0\pm1.3)\keV$. Both
values are consistent with that estimated for ACT-CL J0102--4915
(i.e., $T_{\rm X}=(14.5\pm1.0)\keV$).

The total X-ray luminosities obtained from the mock images in the
$0.5-2.0\keV$ band are $(2.48 \pm 0.03) \times 10^{45}$ (panel (b1))
and $(2.05 \pm 0.03) \times 10^{45} {\,h_{70}^{-2}}\erg\s$
(panel (c1)), respectively, which are similar to the observation of ACT-CL
J0102--4915 (i.e., $(2.19 \pm 0.11) \times 10^{45}{\,h_{70}^{-2}}
\erg\s$; see \citealt{Menanteau2012}).

The total mass of the merging system is $2.0\times10^{15}\dh70\msun$
in fiducial model A, consistent with the old estimate
($(2.16\pm0.32)\times 10^{15}\dh70\msun$) for ACT-CL J0102--4915
by \citet{Menanteau2012}, whereas it is $3.2\times 10^{15}\dh70\msun$
in fiducial model B, consistent with the new estimate
($(3.13\pm0.56)\times 10^{15}\dh70 \msun$) obtained by using the
weak-lensing technique in \citet{Jee2014}. Considering the large
uncertainties in those mass estimates and the possible bias due
to the adoption of the X-ray mass proxies for unrelaxed clusters
\citep[see][]{Nagai2007, Vikhlinin2009}, both fiducial models A
and B can be taken as roughly consistent with the observation in
terms of the cluster mass. To further distinguish the two different
merging scenarios, an accurate mass estimation is required.

We further investigate the temperature obtained from the mock X-ray
emission from each region marked in the left panels of
Figure~\ref{pic:flash_specT} for the two fiducial models.
The resulting temperature distributions against the X-ray emitting
regions are shown in the right panels of Figure~\ref{pic:flash_specT}
for the \textit{Chandra} observation (panel (a2)), fiducial model
A (panel (b2)), and fiducial model B (panel (c2)), respectively. As seen from the
figure, the temperature distributions are roughly consistent with the
observational one obtained for ACT-CL J0102--4915, although the
temperature uncertainties are larger than the observational ones
because of the limited photon numbers in the outer regions resulting
from both models.

We also reproduce the MB model \citep{Molnar2015} for ACT-CL
J0102--4915 ($V=2250\kms,\, P=300\dh70\kpc$) by using the FLASH code,
in order to compare our model results with theirs in detail. The main
results are summarized in Table~\ref{tab:results}. We find
that the X-ray emission resulting from the MB model extends to larger
scales compared with those from fiducial models A and B, mainly due
to a later merging stage adopted in the MB model. The
SZ decrement at the center resulting from the MB model ($-1030\,{\rm
\mu K}$) is also consistent with observations.  However, its mean
temperature and total X-ray luminosity are $(9.9\pm0.6)\keV$ and
$(1.77\pm 0.13) \times 10^{45} {\,h_{70}^{-2}}\erg\s$, respectively,
substantially lower than those from the observations and our models.


\subsubsection{The Mach Number} \label{sec:result:obs:mach}

Figure~\ref{pic:flash_shock1d} shows the hydrodynamical quantities,
i.e., the electron number density ($n_{\rm e}$), temperature ($T$),
pressure ($P_{\rm g}$), and entropy (defined as
$S\equiv k_{\rm B}Tn_{\rm e}^{-2/3}$), across the SE shock
discontinuity (leading the secondary cluster) resulting from 
fiducial models A (left panels) and B (right panels), respectively.
All those profiles are measured along the line across the centers
of the two clusters (in the $z'=0$ plane). The vertical dashed
lines indicate the location of the bow shock. The Rankine--Hugoniot
conditions are adopted to determine the Mach number $\mathcal{M}$
(see eqs.~89.6--89.8 in \citealt{Landau1959}). We estimate
$\mathcal{M}$ from the jump in the temperature profile for its
well-defined discontinuity, and find $\mathcal{M}=4.3$ and $3.8$
for fiducial models A and B, respectively. The expected jumps of other
quantities from the Mach number are also shown as the horizontal
dotted lines in Figure~\ref{pic:flash_shock1d}.
The Mach number from the MB model that we reproduce by using
the FLASH code is $\sim 6.0$.

In the models the NW and SE shocks (moving outward in front of the primary and
secondary clusters, respectively) are coincident in
positions with the observed radio relics of ACT-CL J0102--4915
\citep{Lindner2014}. To compare with the observations,
we also estimate the Mach numbers of the SE and NW shocks from
the spectroscopic-like temperature map (see eq.~6 in
\citet{Mastropietro2008}). They are in general smaller than the
actual values (measured in the $z'=0$ plane) due to the projection
effect \citep[see][]{Mastropietro2008, Machado2013}. The inferred
Mach numbers (SE, NW) for model A, model B, and the MB model
are  $(2.9,\,2.4)$, $(2.7,\,2.5)$, and $(4.6,\,-)$, respectively
(no clear NW shock found in the MB model). The results of our
fiducial models are both consistent with that reported in
\citet{Lindner2014}, where $\mathcal{M}=2.5^{+0.7}_{-0.3}$
is estimated from the spectral index of the NW radio relic
of ACT-CL J0102--4915 (see the comparison between the Mach
numbers derived from the X-ray and the radio observations in
\citet{Akamatsu2013}). A tight constraint on the Mach numbers
derived from the temperature profile across the SE and NW shocks
in the X-ray observation may provide additional information to
distinguish or falsify fiducial models A and B proposed in
this study and the models in \citet{Molnar2015}.

\subsubsection{The Relative Radial Velocity}  \label{sec:result:obs:vel}

\citet{Menanteau2012} estimate the observed relative radial velocity along the
LOS ($586 \pm 96\kms$) between the two components (NW and SE) of ACT-CL
J0102--4915 based on the galaxy redshift distribution, while the observed
relative radial velocity between the NW component and the cluster BCG located
in the SE component is
$731\pm 66\kms$.  The relative radial velocities resulting from fiducial
model A and fiducial model B are $960\kms$ and $1820\kms$ respectively,
which are directly estimated from the peculiar velocities of the NW and the SE
mass centers.  However, it appears that our model results may not be in
contradiction
with the observations, because the values of the relative radial velocities
are obtained from different methods and the observationally determined values
may not reveal the real relative radial velocities of the mass centers.

\begin{figure} 
\centering
\includegraphics[width=0.4\textwidth]{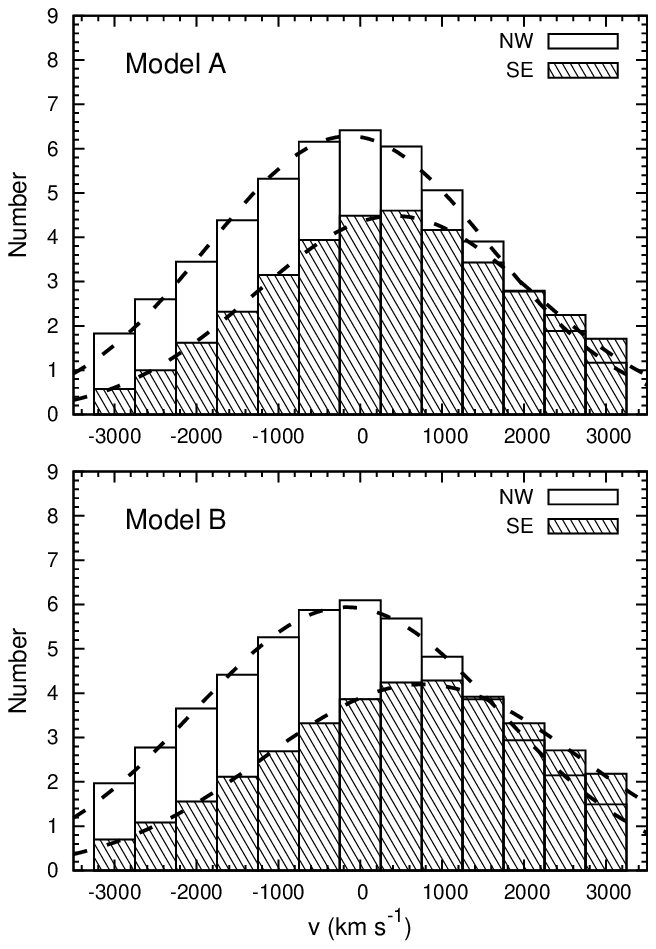}
\caption{Radial velocity distributions along the LOS of the DM
particles in the NW and the SE cluster components for fiducial
model A (top panel) and fiducial model B (bottom panel).
The distributions for the NW and the SE components are normalized
to 51 and 36 in the total number to directly compare with the
observations shown in fig.~9 in \citealt{Menanteau2012}.
The dashed lines give the best-fit Gaussian distributions for the
velocity distributions. The relative radial velocities measured
from the best fits are $560$ and $910\kms$ for fiducial model
A and fiducial model B, respectively. The values are significantly
smaller than the real relative velocity of the mass centers
($960$ and $1820\kms$, respectively) but
generally consistent with the observations, which illustrates that
the observational estimates may be significantly biased and underestimate the
relative radial velocity between the NW and SE cluster components.
See Section~\ref{sec:result:obs:vel}.
}
\label{pic:flash_velocity}
\end{figure}

\begin{figure*} 
\centering
\includegraphics[width=0.8\textwidth]{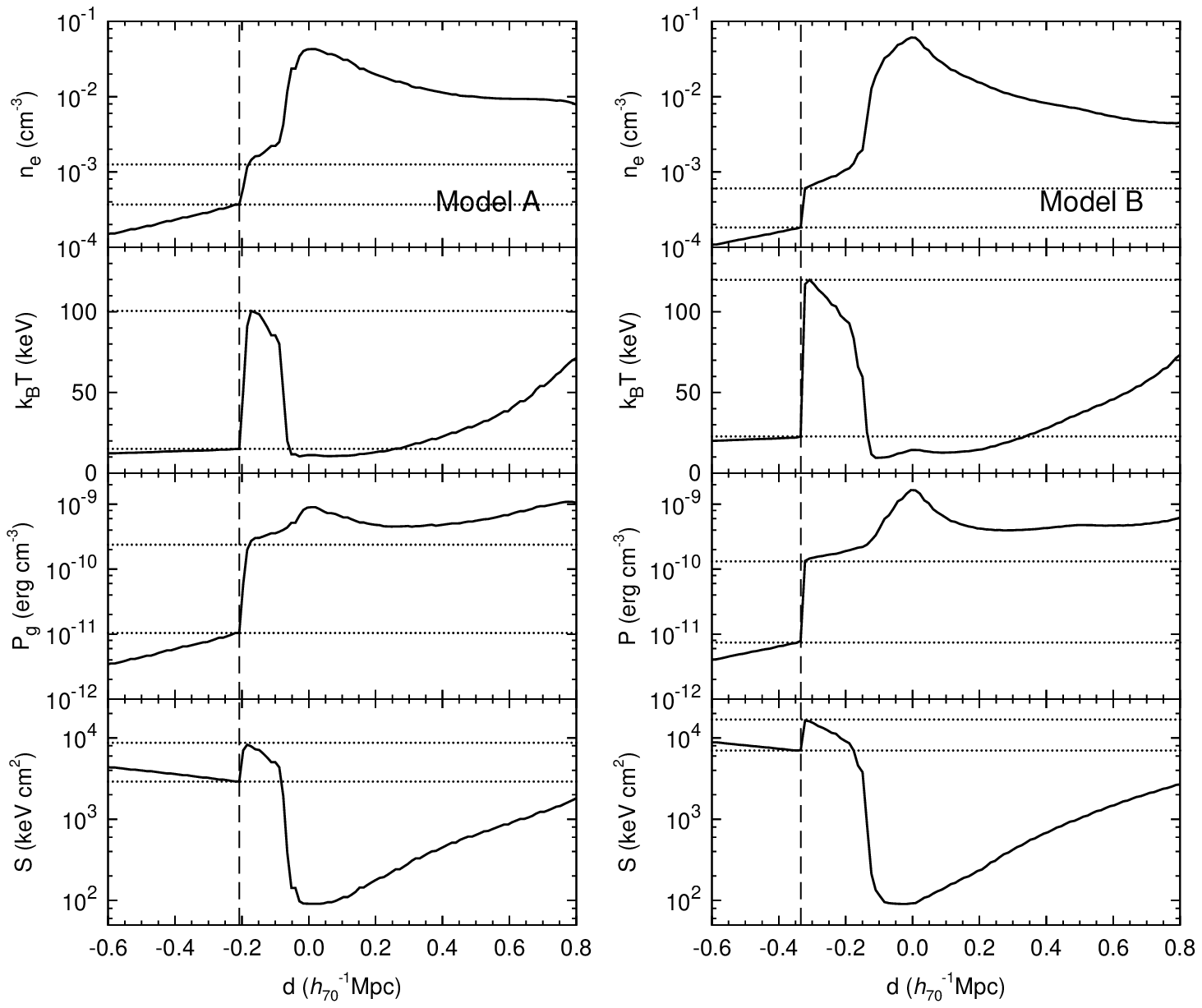}
\caption{Electron number density, gas temperature, pressure, and
entropy profiles across the SE shock discontinuity resulting from 
fiducial model A (left panels) and fiducial model B (right
panels). The vertical dashed lines mark the location of the bow shock.
The center of the secondary cluster is at $d=0$. The Mach number
${\mathcal{M}}$ determined from the discontinuity in the temperature
profile is $\mathcal{M}=4.3$ and $3.8$ for fiducial models A and B,
respectively. The dotted horizontal lines give the expected jump of
the quantities from the obtained Mach number.
See Section~\ref{sec:result:obs:mach}.
}
\label{pic:flash_shock1d}
\end{figure*}

The values of relative radial velocities obtained from different methods may
differ from each other significantly (e.g., by a factor of 2).  To illustrate this point, we
model the velocity distributions for the DM particles in the NW and the SE
cluster components of fiducial models A and B in
Figure~\ref{pic:flash_velocity}, where the DM particles separated from the NW
or the SE mass centers within a projected distance of $400\dh70\kpc$ on the
plane of the sky are referred to as the NW or the SE cluster component. The
relative radial velocities measured from the best-fit Gaussian distributions
for the NW and the SE velocity distributions (see dashed lines in the figure)
are $560$ and $910\kms$ for fiducial models A and B, respectively,
which are significantly lower than the real ones, but generally consistent with
the observational values. The reason for the low values obtained in
Figure~\ref{pic:flash_velocity} is that the DM particles (or the
galaxies in the observations) are grouped into two subsets by their projective
distances to the NW and the SE mass centers, respectively; and the merging
process has destroyed the boundaries of the original clusters, so that the SE
(NW) subset contains the particles or galaxies initially belonging to the NW
(SE) cluster (whose velocities, however, are still close to their original host).
Figure 9 indicates that the measurement used in Menanteau et al.\ (2012) may
underestimate the relative velocity between the two clusters significantly.
Note that the degree of the underestimate depends on the overlapping fraction
of the two clusters along the LOS. For example, if the two clusters
are at a relatively later merging stage after the pericentric passage so that
they have a larger separation and less overlapped, the underestimate may be
significantly smaller than the factor of 2.

In addition to the above significant factor, the relative radial velocity
estimations may be also affected by a few other factors.
(a) The velocities inferred from different components (e.g., DM, galaxies)
within a cluster are different \citep{Dolag2013}. For ACT-CL J0102--4915, the
observed relative radial velocity is estimated from the motions of galaxies,
while the velocity resulting from the model is based on the motions of DM
particles.  This difference could introduce an error on the order of $100\kms$.
(b) The relative radial velocity resulting from a model is also sensitive to
the choice of the projection angle $\alpha$ and $i$.  If choosing $i=15\arcdeg$
rather than $30\arcdeg$ in fiducial model B (the X-ray morphology does not
differ significantly), the relative radial velocity decreases by a factor of
$\sim$2.

\subsection{Discussion on the Effects of the Gas Fraction Profile}
\label{sec:result:fgas}

\begin{figure} 
\centering
\includegraphics[width=0.45\textwidth]{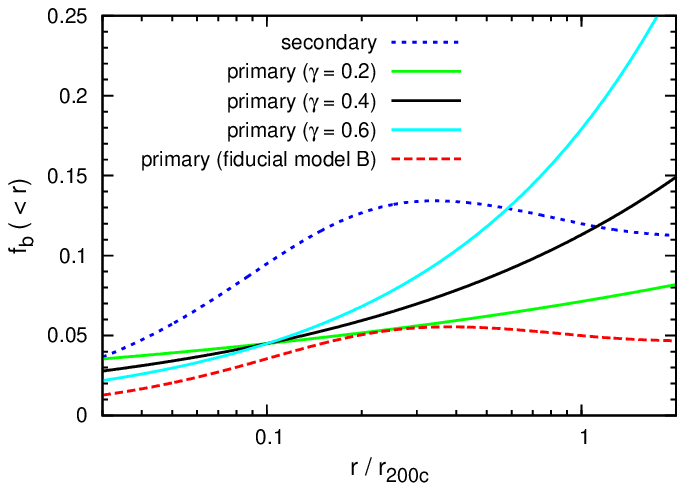}
\caption{Cumulative gas fraction profiles for the primary and the secondary clusters.
The green, black, and cyan solid lines represent the profiles of the primary clusters
following Equation~(\ref{eq:fgas}) with $A=0.045$ and $\gamma=0.2,\,0.4$, and 0.6,
respectively. The blue dotted line represents the profile of
the secondary cluster in the extended case B mergers. The red dashed line represents
the profile of the primary cluster in fiducial model B.
In the central region, the profiles shown by the solid lines are close to
that of fiducial model B; but at the radius $r_{200}$, the gas fractions shown
by the solid lines are all higher than $0.05$. See Section~\ref{sec:result:fgas}.
}
\label{pic:modelB_e_profile}
\end{figure}

According to the simulations, fiducial model B could reproduce most of
the observational features of ACT-CL J0102--4915, but with two deficiencies:
(1) less X-ray emission is produced in the outer region of the merging
cluster compared with the observations (see Section~\ref{sec:result:obs:xray});
(2) the adopted gas fraction of the primary cluster ($0.05$) is substantially
lower than the typical value of massive galaxy clusters ($\sim0.13$). The
simplified gas density distribution for the galaxy clusters in the simulation
results in an approximately flat cumulative gas fraction profile while
$r>0.2r_{\rm 200}$, which, however, does not well represent the situations in
the observations and cosmological simulations \citep{Battaglia2013,Mantz2014}.
We find that setting the cumulative gas fraction as a function of the radius
following the constraints from the observations may solve the above discrepancies.
We further perform simulations with different gas density profiles described below,
denoted as the extended case B mergers, by using the GADGET-2 code. For the
primary cluster, we assume that the cumulative gas fraction profile follows a
power-law form,
\be
f_{\rm b1}(<r)=A\left(\frac{r}{r_{\rm f}}\right)^{\gamma},
\label{eq:fgas}
\ee
where $r_{\rm f}\equiv0.1r_{\rm 200}$ is the scale radius. The gas
density distribution can then be numerically determined from the enclosed
DM mass distribution. For the secondary cluster, the gas density distribution
still follows the Burkert profile (see Eq.~\ref{eq:rhogas1}) as that in 
fiducial model B, but with two differences, i.e., $f_{\rm b2}=0.12$ and
$r_{\rm c}=r_{\rm s}/2$. We find that the simulation results with the
relatively large gas core size of the secondary cluster ($\sim100\dh70\kpc$)
give a good match to ACT-CL J0102--4915. As an example, we show the different
cumulative gas fraction profiles adopted in the extended case B mergers when
$A$ is set to $0.045$ in Figure~\ref{pic:modelB_e_profile}. In the central
region ($\sim0.05-0.3r_{\rm 200}$), the gas fractions in Equation~(\ref{eq:fgas}) are close to $0.05$,
similar to that of fiducial model B; at the radius $r_{200}$, the gas
fractions are $0.07,\,0.11,\,0.18$ for those cases with $\gamma=0.2,\,0.4,\,0.6$,
respectively, generally consistent with the observational constraints from the
observations \citep{Mantz2014}. It is worth noting that the power-law form
for the cumulative gas fraction profile is unrealistic for the outer region
of the galaxy clusters when the gas fraction is much higher than the
cosmological average value (e.g., $r>2r_{\rm 200}$ for the $\gamma=0.4$ case),
which, however, has little effect on our results since the DM density distribution
drops significantly outside $r_{\rm 200}$ in our models.
The merger configuration of the extended case B mergers follows that of
fiducial model B with the parameter set $(M_1,\, \xi,\, V,\, P) = (2.5\times
10^{15}\dh70\msun,\, 3.6,\, 2500\kms,\, 800\dh70\kpc)$.

\begin{figure*} 
\centering
\includegraphics[width=1.0\textwidth]{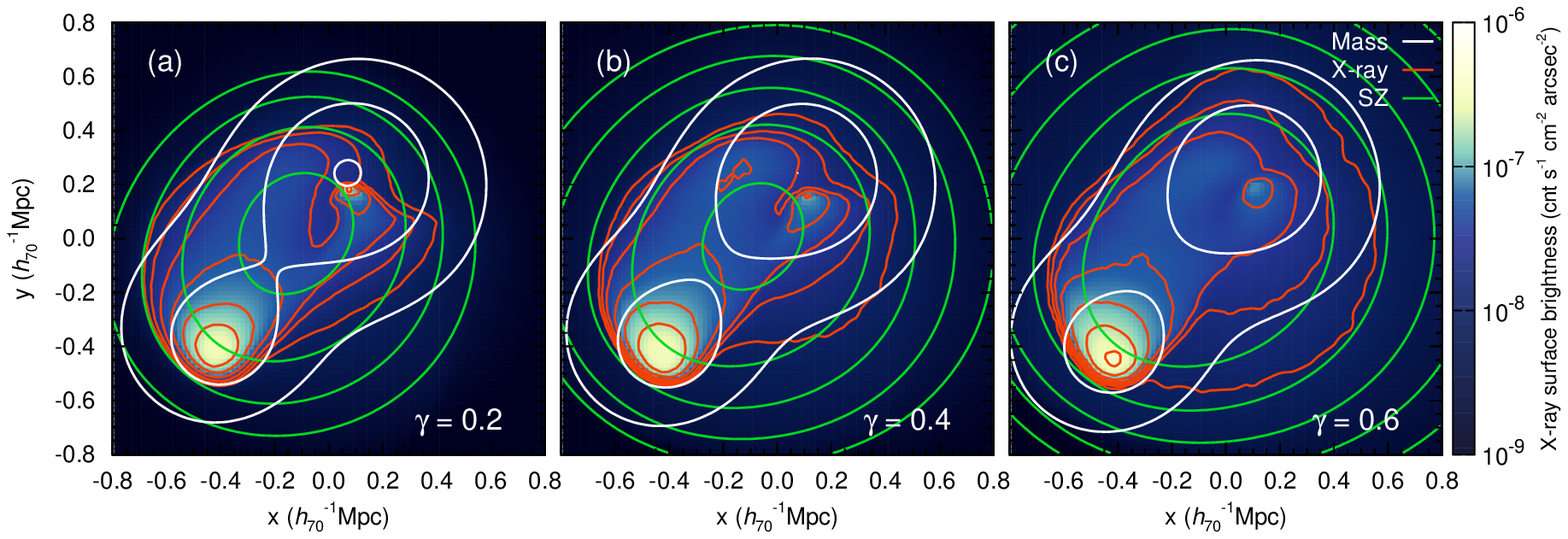}
\caption{Similar to those in Figure~\ref{pic:modelA_inclination},
but for the simulated merging clusters in the extended case B mergers (where the cumulative gas fraction profile follows Equation~(\ref{eq:fgas})). Panels (a),
(b), (c) show the results for the cases with $\gamma=0.2,\, 0.4$, and 0.6,
respectively, and $A$ is fixed to $0.045$. The viewing direction is set to be
the same with $(\alpha,\, i) =(-90\arcdeg,\, 40\arcdeg)$ for all the three
cases. The merging time for the three snapshots is $t=0.17\Gyr$. The simulation results of panel
(b) are similar to those of fiducial model B, but with stronger X-ray
emission in the outer region of the merging cluster. See
Section~\ref{sec:result:fgas}.
}
\label{pic:modelB_e_fgas}
\end{figure*}

Figure~\ref{pic:modelB_e_fgas} shows the simulation results for the different
cumulative gas fraction profiles (i.e., $\gamma=0.2,\,0.4$ and 0.6), which
correspond to the solid lines in Figure~\ref{pic:modelB_e_profile}, respectively.
As seen from Figure~\ref{pic:modelB_e_fgas}, the simulation results
presented in panel (b) are quite similar to those of fiducial model B (see
Fig.~\ref{pic:modelB_inclination}b), but the X-ray emission in the outer
region of the merging cluster increases as the gas fraction at the virial
radius becomes higher. Panel (a) shows a highly asymmetric twin-tailed
structure in the X-ray image and a remarkable secondary X-ray peak close
to the center of the primary cluster, which do not match the observations
well. Unlike that in panel (b), the secondary X-ray peak in panel (a) is
still clear in its mock \textit{Chandra} X-ray image. Panel (c) also fails
to reproduce the observations since no clear wake-like structure appears in
the X-ray image. Furthermore, the different cumulative gas fraction profiles
with $A=0.03$ and $0.06$ are also tested. We find that $A=0.045$ is preferred to
match the X-ray luminosity of ACT-CL J0102--4915.

We further re-run a FLASH simulation for the extended case B merger with
$A=0.045,\, \gamma=0.4$ (denoted as extended model B) and compare
the model with the observations as done in Section~\ref{sec:result:obs}.
The main results are summarized in Table~\ref{tab:results} (see also
Figs.~\ref{pic:flash_specT}d1 and \ref{pic:flash_specT}d2). We find the
following points for extended model B. (1) The mock X-ray image and
the temperature distribution are similar to that of fiducial model B
(see Figs.~\ref{pic:flash_specT}c1 and \ref{pic:flash_specT}c2). But the X-ray
emission in the outer region obtained from extended model B is more
significant, which is comparable with the observational one.
(2) The total X-ray luminosity in the $0.5-2.0\keV$ band and the
best-fitted X-ray temperature are $(2.08\pm0.03)\times10^{45}  {\,h_{70}^{-2}}\erg\s$ and $(18.0\pm1.8)\keV$, respectively.
The Mach number across the SE shock (measured in the $z'=0$ plane) is $3.1$,
and the relative radial velocity between the two clusters is $2060\kms$. The
quantities above are all consistent with those of fiducial model B except
that the temperature is about $20\%$ higher. (3) The central temperature
decrement of the SZ effect of extended model B is, however,
$-1850{\,\mu{\rm K}}$, much lower than the measurement of ACT-CL J0102--4915.
The non-thermal pressure may play a non-negligible role in this situation;
and ignoring the non-thermal pressure in our models
may lead to the overestimation of the central temperature decrement by
dozens of percent (see \citealt{Battaglia2012,LKN09}).

\section{Conclusion}
\label{sec:conclusion}

In this work, we perform a series of simulations of mergers of two
galaxy clusters in order to investigate the merging scenario and
identify the initial conditions for ACT-CL J0102--4915. By surveying
over the space of those parameters that define the merger
configuration, including the mass of the primary cluster, mass ratio,
gas fraction, initial relative velocity, and impact parameter, we
discuss two types of the merger configuration that may be able to
reproduce the observations of ACT-CL J0102--4915, respectively. The
first type is a nearly head-on merger with impact parameter
$\sim 300\dh70\kpc$ (fiducial model A) and the second one is a
highly off-axis merger with $\sim 800 \dh70 \kpc$ (fiducial model B).
The detailed comparison of our simulation result with the observations
of ACT-CL J0102--4915 is summarized in Table~\ref{tab:results}.

Fiducial model A is for an energetic collision of two clusters.
In this model, the central gas core of the primary cluster is completely
destroyed after the first core-core collision. The morphology of the
X-ray surface brightness, the X-ray luminosity, and the temperature
distributions of ACT-CL J0102--4915 can be reproduced, but no wake-like
substructure trailing the secondary cluster is produced by the model.
According to the simulations, fiducial model A is the
merging configuration that best matches the observations when the
impact parameter is smaller than $500\dh70 \kpc$.

Fiducial model B is for a less energetic collision, compared
with fiducial model A, as its impact parameter is much larger.
A remarkable wake-like feature is clearly seen trailing after the
secondary cluster, which is similar to that of ACT-CL J0102--4915.
The total X-ray luminosity of a merging cluster is positively
correlated with the total mass of the system, which may provide a
strong constraint to the model. The total mass in fiducial
model B is $3.2 \times 10^{15} \dh70\msun$, consistent
with the constraint from the weak-lensing technique \citep{Jee2014}.
Our simulation results support that the NW subcluster of ACT-CL
J0102--4915 is more massive than the SE one, in agreement with the
measurement by \citet{Jee2014} but not that by \citet{Zitrin2013}.
The mass ratio between the NW and SE components of the cluster is
$3.6$ for fiducial model B, which is somewhat higher than the
estimate in \citet{Jee2014}.

Fiducial model B can reproduce most of the basic features of ACT-CL
J0102--4915, but it produces less X-ray emission in the outer region of
the merging cluster compared to the observations. The reason might be
that the adopted gas density profile in the model may not well represent
the reality. Adopting the cumulative gas fraction as a function
of the radius ($\sim0.05$ at $0.1r_{\rm 200}$ and $\sim0.11$ at
$r_{\rm 200}$) motivated by observations \citep{Mantz2014} can
solve this discrepancy; and it may also provide a natural explanation to
the low gas fraction ($0.05$) of the primary cluster assumed in
fiducial model B. Compared with the models proposed in
\citet{Molnar2015}, fiducial model B presented in this paper
appears to provide a better match to the X-ray morphology and the
best-fit X-ray luminosity and temperature.

In this paper, the initial relative velocity of the two progenitor
clusters of ACT-CL J0102--4915 is high, as suggested by fiducial
model B ($\sim 2500\kms$). The requirement of a high initial relative
velocity for ACT-CL J0102--4915 may enhance the tension between the
rarity of the high velocity mergers of clusters in cosmological
simulations and the existence of the Bullet Cluster
\citep[e.g.,][but \citealt{Thompson2015}]{Lee2010,Thompson2012},
especially when considering the uncommon high mass (see more
discussions  in \citet{Jee2014}) and the small mass ratio of
ACT-CL J0102--4915. According to fiducial model B, the highly
off-axis merger configuration of ACT-CL J0102--4915 is different
from that
of the Bullet Cluster. Nevertheless, as ACT-CL J0102--4915 is
extremely massive and rare at $z\sim 0.87$, the requirement of an
extremely high initial relative velocity may present an even more
significant challenge to our understanding of the structure
formation compared to that by the Bullet cluster.

We thank David Spergel for bringing the galaxy cluster ``El Gordo'' to
our attention.
This research was supported in part by the National Natural Science
Foundation of China under nos.\ 11273004, 11373031, and the Strategic
Priority Research Program ``The Emergence of Cosmological Structures'' of the Chinese Academy of Sciences, Grant No. XDB09000000.  The
software used in this work was developed in part by the DOE NASA
ASC- and NSF supported Flash Center for Computational Science at the
University of Chicago. This research has made use of data obtained
from the \textit{Chandra Data Archive} and the software provided by
the \textit{Chandra X-ray Center} (\textit{CXC}) in the application
package CIAO.

\end{document}